\RequirePackage{rotating}
\documentclass[usenatbib]{mnras}
\usepackage{epsfig}
\usepackage{amsmath}
\usepackage{graphicx}
\usepackage{array}
\usepackage{textcomp}
\usepackage{amssymb}
\usepackage{rotating}
\usepackage{caption}
\usepackage{subcaption}
\def\mnras {{\sc MNRAS}}
\def\apj {ApJ}
\def\apjs {ApJS}
\def\apjl {ApJL}
\def\aj {AJ}

\def\pasp {PASP}
\def\aap {A\&A} 
\def\nat {Nature}
\def\apss{Ap\&SS}

\title[IFS of Massive Void Galaxies]
      {The Many Assembly Histories of Massive Void Galaxies as Revealed by Integral Field Spectroscopy}
\author[A.\ Fraser-McKelvie et al.]
       {Amelia Fraser-McKelvie$^{1,2,3}$\thanks{amelia.mckelvie@monash.edu}, Kevin A.\ Pimbblet$^{1,2,3}$, Samantha J. Penny$^{4}$ and \newauthor Michael J.\ I.\ Brown$^{1,2}$.
        \vspace*{1mm}\\
        $^{1}$ School of Physics and Astronomy, Monash University, Clayton, Victoria 3800, Australia\\
        $^{2}$ Monash Centre for Astrophysics (MoCA), Monash University, Clayton, Victoria 3800, Australia\\
	$^{3}$ E. A. Milne Centre for Astrophysics, Department of Physics and Mathematics, University of Hull, Cottingham Road, \\Kingston-upon-Hull, HU6 7RX, UK\\
	$^{4}$ Institute of Cosmology and Gravitation, University of Portsmouth, Dennis Sciama Building, Burnaby Road, Portsmouth, PO1 3FX, UK\\
	}

\begin{document}
\maketitle
\begin{abstract}
We present the first detailed integral field spectroscopy study of nine central void galaxies with $\textrm{M}_{\star}>10^{10}~\textrm{M}_{\odot}$ using the Wide Field Spectrograph (WiFeS) to determine how a range of assembly histories manifest themselves in the current day Universe. While the majority of these galaxies are evolving secularly, we find a range of morphologies, merger histories and stellar population distributions, though similarly low H$\alpha$-derived star formation rates ($<1~\textrm{M}_{\odot}~\textrm{yr}^{-1}$). Two of our nine galaxies host AGNs, and two have kinematic disruptions to their gas that are not seen in their stellar component.
Most massive void galaxies are red and discy, which we attribute to a lack of major mergers. 
Some have disturbed morphologies and may be in the process of evolving to early-type thanks to ongoing minor mergers at present times, likely fed by tendrils leading off filaments. The diversity in our small galaxy sample, despite being of similar mass and environment means that these galaxies are still assembling at present day, with minor mergers playing an important role in their evolution. 

We compare our sample to a mass and magnitude-matched sample of field galaxies, using data from the Sydney-AAO Multi-object Integral field spectrograph (SAMI) galaxy survey. We find that despite environmental differences, galaxies of mass $\textrm{M}_{\star}>10^{10}~\textrm{M}_{\odot}$ have similarly low star formation rates ($<3~\textrm{M}_{\odot}~\textrm{yr}^{-1}$). The lack of distinction between the star formation rates of the void and field environments points to quenching of massive galaxies being a largely mass-related effect.
\end{abstract}
 
  
\section{Introduction}
Galaxy evolutionary mechanisms can broadly be split into two categories: environmentally driven and mass driven effects.
The former category include processes such as galaxy harassment \citep[e.g.,][]{Moore96}, ram pressure stripping \citep{Gunn72}, viral shock heating \citep[e.g.,][]{Dekel06,Cen11} and mergers \citep{Toomre72}, whilst the latter include secular processes such as AGN feedback \citep[e.g.,][]{Croton06} and mass quenching \citep[e.g.,][]{Kauffmann03,Geha12}. Generally it is difficult to separate these processes throughout a galaxy's lifetime, especially in high-density environments where many of these mechanisms act on a galaxy simultaneously.

Examining the low-density regions of the Universe offers an unparalleled view into galaxy evolution in isolation. Filling the space between filaments and clusters, cosmic voids are spatially the largest regions of the Universe \citep[e.g.,][]{AragonCalvo10}, with the lowest mean galaxy density \citep[10-20 per cent of the mean cosmic density;][]{Pan12}. Typical void regions are roughly spherically symmetric, or prolate spheroids \citep{Platen08}, with mean effective radii 17 $h^{-1}$ Mpc \citep{Pan12}. 

Wide field spectroscopic surveys such as the Sloan Digital Sky Survey \citep[SDSS;][]{Abazajian09} delineate the large scale structure of the cosmic web and allow the properties of the void galaxy population to be studied in detail. Observationally, \citet{Pan12} created a catalogue of voids from the SDSS Data Release 7 (DR7) based on a nearest neighbour technique that was employed by \citet{Hoyle04} for the 2dF Galaxy Redshift Survey. These studies confirm that the few galaxies located within voids typically lie close to the edges, leaving galaxies in central regions extremely isolated.

The low density of cosmic void regions allow galaxies to evolve in relative solitude, removing many environmental effects.
It is for this reason voids are excellent laboratories to study the secular effects of galaxy and halo mass on galaxy evolution.
Many studies have found a strong dependence on environment for several galaxy properties, particularly in voids. 
When compared to galaxies in denser environments, void galaxies are thought to posses higher specific star formation rates \citep[e.g.,][]{Grogin00, Park07, von08, Ricciardelli14, Liu15}, and are on average bluer and discier than similarly luminous galaxies in denser environments \citep[e.g., ][]{Rojas04, Rojas05, Hoyle05, Deng08, Hoyle12, Moorman15}. 
These characteristics are generally attributed to a limited opportunity for interactions due to their relative isolation, leaving more gas available for star formation. \citet{Moorman16} finds void regions are particularly nurturing for dwarf galaxies.

While a consistent picture of a bluer void galaxy population has been presented above, recent work instead finds very little difference between galaxies in voids and those in structure, especially when a multi-wavelength approach is taken \citep[e.g.,][]{Beygu16}. \citet{Penny15} show no difference in the mid-infrared (IR) colour of 592 galaxies located in voids and an analogous sample of field galaxies; a wide spread in recent star forming activity is found throughout both populations. Optically red void galaxies display a range of mid-IR colour, suggesting dust-obscured star formation or star formation occurring on the outskirts of the galaxies, missed by single fibre spectroscopy. This range of mid-IR colour suggests a variety of assembly histories must exist in the void galaxy population.
This finding has been backed up by smaller, multi-wavelength surveys such as \citet{Sage97}, who detect CO emission from four Bootes void galaxies that is comparable to that seen from interacting field galaxies. \citet{Beygu13} and \citet{Das15} also detect molecular gas in void regions, inferring star formation rates and dynamical activity similar to that of galaxies in denser environments.
Detailed, spatial studies of these galaxies are needed to locate star forming regions within void galaxies, the extent of this star formation, and the mechanisms by which it is occurring.

\begin{figure}
 \centerline{ 
\epsfig{file=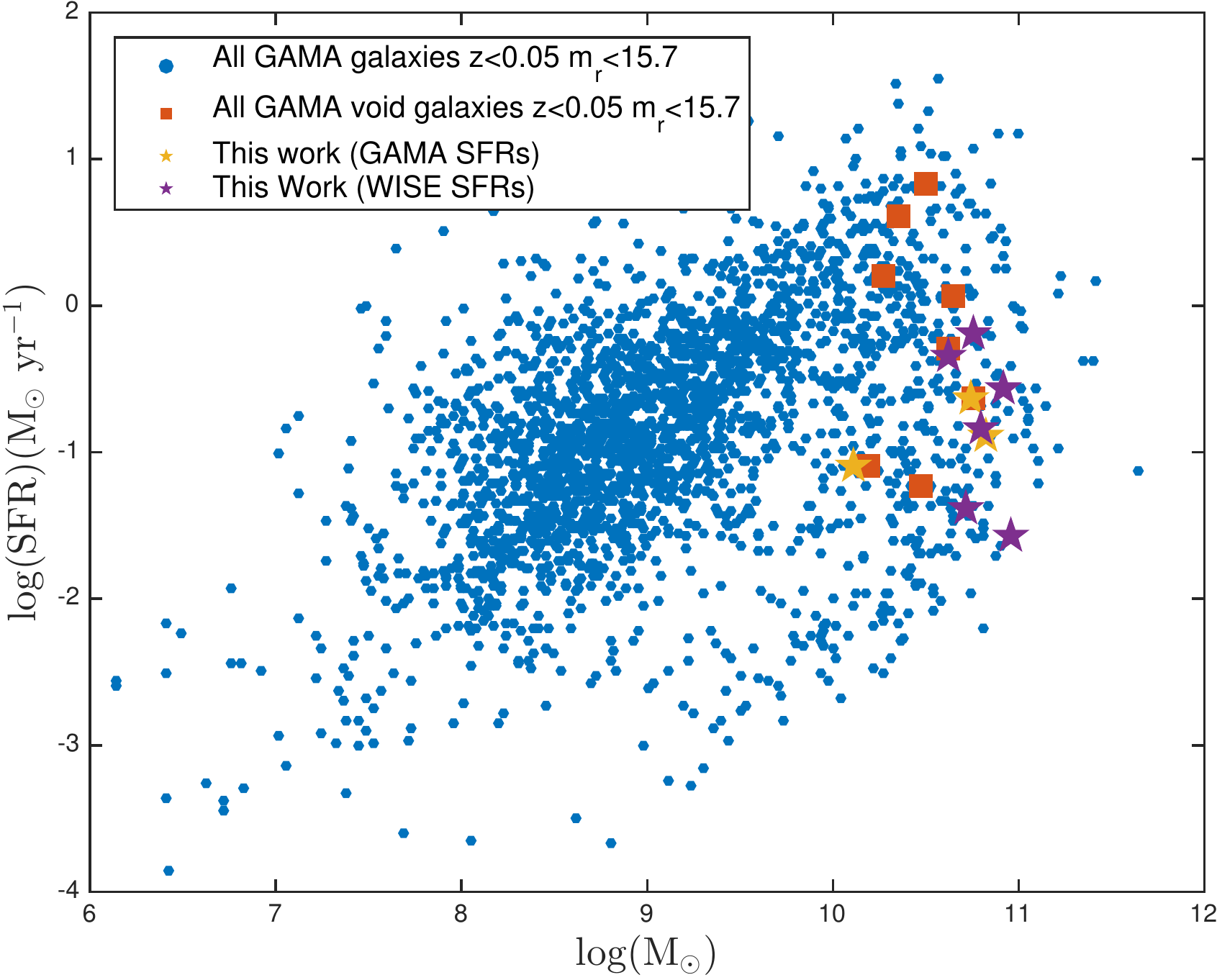,angle=0,width=3.5in}
}
\caption{The H$\alpha$-derived star forming main sequence of all GAMA galaxies with $z<0.05$. Blue points are all GAMA galaxies, orange are isolated and central void galaxies with $\textrm{M}>10^{10} \textrm{M}_{\odot}$ from the catalogue of \citet{Penny15}. Yellow points are GAMA galaxies that were observed for this work. Purple points were also observed for this work, but GAMA SFRs were not available, so WISE $W3$ 12$\mu$m-derived SFRs using the relation of \citet{Cluver14} were used instead. 
The isolated and central galaxies within voids cover a wide range of star formation rates, extending through both the red sequence and blue cloud, suggesting multiple assembly histories of these galaxies.} 
\label{SFMS}
\end{figure}

Massive void galaxies constitute an unique population as their high luminosity permits detailed studies into galaxy kinematics and merger histories over relatively short integration times.
The most massive void galaxies are the central galaxy of their dark matter halo and have masses $\textrm{M}_{\star}>10^{10} \textrm{M}_{\odot}$, similar to that of brightest group galaxies or small brightest cluster galaxies \cite[BCGs; e.g.,][]{Oliva14}.  Their wide spread of current day star formation, as shown in Figure~\ref{SFMS}, means that there must be a mixture of mass growth processes at play over a range of time scales. 
Simulations predict that mass is accumulated through continued star formation fed by filaments of gas accreting onto both isolated and central group void galaxies \citep[e.g.,][]{Kreckel12, Beygu13}. This is confirmed observationally by \citet{Penny15}, who find the dominant morphology of high mass void galaxies is discy with many red spirals \citep[e.g.,][]{Masters10, Bonne15}.

At masses greater than $\textrm{M}_{\star}>10^{10} \textrm{M}_{\odot}$ the mass assembly histories of galaxies are increasingly dominated by major mergers \citep{Kauffmann03, Baldry04}, creating early-type galaxies that have quenched by current times. This apparent disparity in formation history can be resolved by spatial mapping of these massive void galaxies. Are these galaxies evolving secularly as their isolated nature would suggest, or have enough interactions occurred that they have morphologically (and kinematically) transformed?

Single fibre spectroscopy often catches only old stellar populations at the heart of massive galaxies, misdiagnosing them as passive, though mid-IR photometry suggests these galaxies are not as quenched as they appear \citep[e.g.,][]{Penny15}.  In a wide-field multi-object redshift survey, each galaxy is targeted with just one fibre. As we show in Section~\ref{Motivation}, for massive low redshift galaxies, this fibre will only capture light from the central region of a galaxy, potentially missing a star forming disc. 

Integral field spectroscopy (IFS) allows the features of a galaxy to be spatially mapped by recording many spectra of the same object simultaneously. Properties such as stellar and gas kinematics, along with areas of star formation and emission line regions can be mapped out to several effective radii, conveying a comprehensive, spatially resolved picture of a galaxy. Traditional single object IFS units are able to image one galaxy in extreme detail, though are observationally expensive. 
The advent of multi-object IFS surveys such as the SAMI galaxy survey \citep{Croom12}, and Mapping Nearby Galaxies with APO \cite[MaNGA;][]{Bundy15} will allow galaxy populations to be examined spatially with number statistics like never before. 
These surveys are currently in their infancy, and will take several more years to build up statistically significant samples targeting specific galaxy sub-populations. It may take some time for a population of rare objects such as massive, isolated void galaxies to be observed. 

We exploit this niche by targeting the small sub-population of massive void central galaxies using the Wide Field Spectrograph \citep[WiFeS;][]{Dopita07,Dopita10} single-object IFS instrument for a detailed study of their properties.
In this study we aim to identify and classify any morphological, kinematic or spectroscopic similarities in massive central void galaxies and compare them to a mass-matched sample of field galaxies. 

We present IFS observations of nine massive isolated and central void galaxies along with a comparison sample of nine field galaxies to determine the relative contribution of environmental and mass effects to their current day appearance. The paper is organised as follows: in Section~\ref{targets} we define the GAMA massive isolated and central void galaxy population and analyse any bias in our target selection sample. In Section~\ref{observations} we describe the observations and data reduction, and in Section~\ref{stellar} we present the stellar and gas kinematic derivations. Section~\ref{SFRsection} describes the star formation rates and in Sections~\ref{results} \& \ref{ComparisonSection} we present our results along with a comparison to a mass and magnitude-matched sample of field galaxies from the SAMI galaxy survey. Throughout this paper we use a flat $\Lambda$CDM cosmology with $\Omega_{\Lambda}=0.7$, $\Omega_{M}=0.3$, and $H_{0}=70 \textrm{km}~\textrm{s}^{-1}$. A \citet{Salpeter55} initial mass function (IMF) is used for all star formation rates and stellar masses presented in this work unless stated otherwise.

 \section{Void galaxy target selection}
 \label{targets}
 We draw our central and isolated massive void galaxy targets from the Galaxy And Mass Assembly \citep[GAMA;][]{Driver11} survey void catalogue of \citet{Penny15}. The catalogue identifies all galaxies within the GAMA equatorial survey regions located in the SDSS voids defined by \citet{Pan12} using the Void Finder algorithm out to $z\sim0.1$. 
 
 The Void Finder algorithm is described in \citet{Hoyle02} and uses a nearest neighbour approach to find statistically significant cosmic voids in volume limited wide-field redshift survey data. The method takes galaxies located in field or wall regions and computes the third nearest neighbour distance, $d_3$, for each galaxy. All galaxies with $d_{3}>6.3 h^{-1} \textrm{Mpc}$ are considered as potential void galaxies and removed from the sample. What remains are potential wall galaxies, which are mapped and empty spheres grown between the wall regions. If a sphere can be created with radius $> 10 h^{-1} \textrm{Mpc}$, the region is considered a void, and any galaxies located within this void region are void galaxies. Non-spherical voids are created by the addition of two neighbouring regions if there is sufficient overlap. This method has proven robust on both observational and simulated galaxies and dark matter. Average void density was found to be $<20\%$ of the cosmic mean at void edges, dropping to $\sim10\%$ in central regions.

\begin{figure}
\centerline{
\epsfig{file=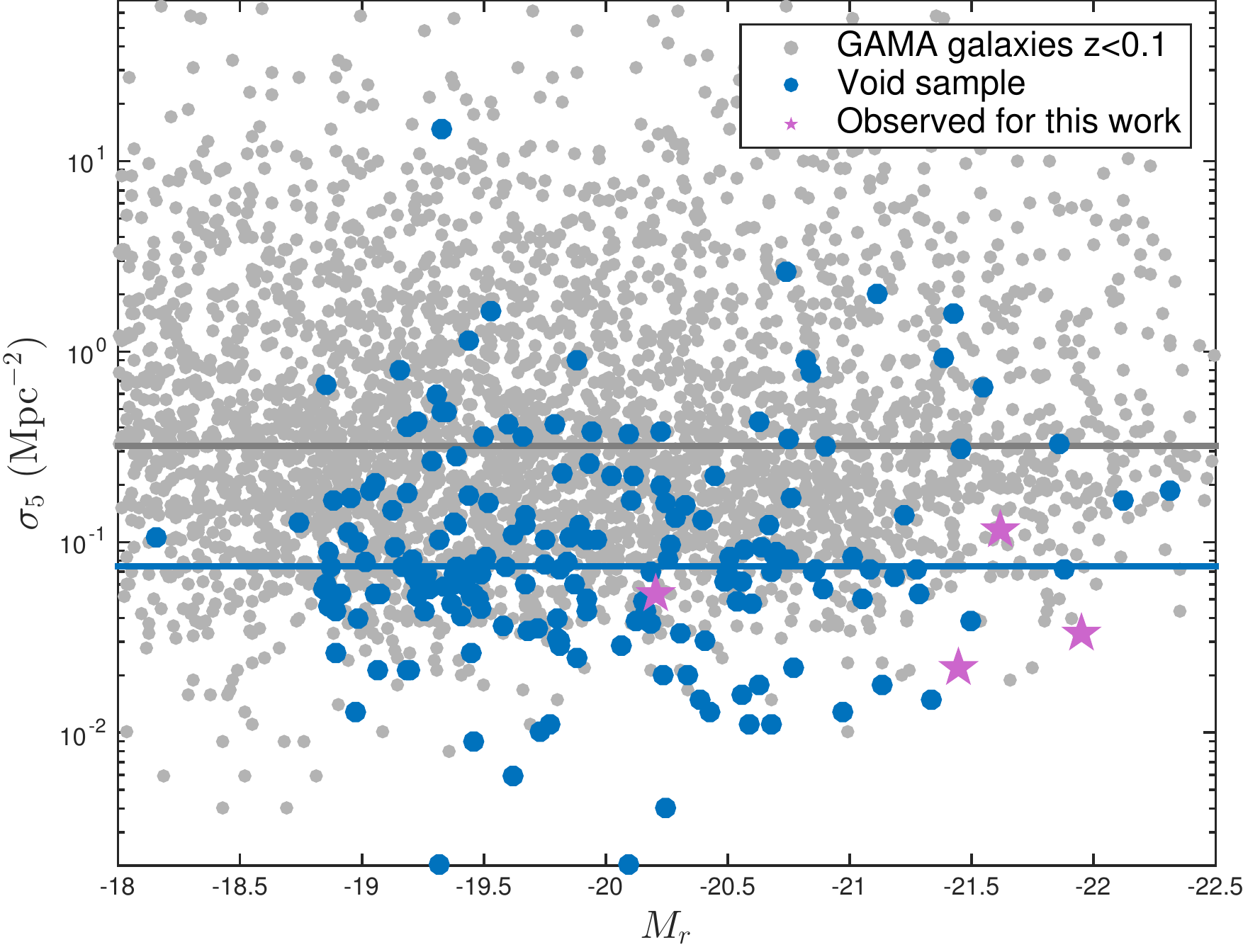, angle=0, width=3.5in}
}
\caption{The $5^{\textrm{th}}$ nearest neighbour surface density, $\sigma_{5}$, for the galaxies in the \citet{Penny15} void galaxy catalogue with matches to the environmental measures catalogue of \citet{Brough13} (blue points, with a blue median line of 0.0745 $\textrm{Mpc}^{-2}$). All GAMA galaxies with $z<0.1$ in this catalogue are shown as grey points with a grey median line value of 0.319 $\textrm{Mpc}^{-2}$, and those observed for this work in the GAMA catalogue are shown as pink stars. On average, the void galaxies in our sample (and all those observed) are located in regions of lower environmental density than galaxies in all environments in GAMA.  }
\label{envplot}
\end{figure}

 While the \citet{Pan12} catalogue contains only galaxies with magnitudes brighter than $M_{r}=-20.09$, \citet{Penny15} include all GAMA galaxies brighter than $M_{r}=-18.4$. This fainter magnitude limit includes more low-mass and low-surface brightness galaxies, and also decreases the size of the void regions as fainter, less-clustered galaxies extend further into void regions from surrounding filaments. For this reason, only galaxies out to 0.75 $R_{void}$ are included, ensuring isolation from large-scale structure. This gives us confidence that our void galaxies are indeed located in bona fide void regions, and not surrounding filamentary regions.
 We check our void galaxies are in sufficiently underdense regions of the cosmic web by finding the $5^{\textrm{th}}$ nearest neighbour distance ($d_{5}$) and surface density ($\sigma_{5}$) of all galaxies in the void sample using the \citet{Brough13} environmental measures catalogue, as shown in Figure~\ref{envplot}. For void galaxies within the GAMA G15 field, we find a median environmental density of $\sigma_{5}=0.0745~\textrm{Mpc}^{-2}$, compared to $\sigma_{5}=0.319~\textrm{Mpc}^{-2}$ for all galaxies in GAMA with $z<0.1$. We confirm that the galaxies in our void sample are located in under-dense regions of the Universe.

 We make a further mass cut of $\log(\textrm{M}_{\star}) > 10^{10} \textrm{M}_{\odot}$ and use the GAMA galaxy group catalogue of \citet{Robotham11} to select only isolated and central group galaxies in our sample and remove interlopers. Only galaxies with publicly available GAMA SFRs \citep{Hopkins13} and stellar masses \citep{Taylor11} were included in this work. 
 
 The final sample of massive void galaxies comprises 76 isolated void galaxies, and 37 central group galaxies residing in groups of five members or less, all with $z<0.1$. These galaxies are all the most massive objects in their dark matter haloes. We draw our observing targets from this sample, and supplement with SDSS galaxies of the same constraints outside of the GAMA survey right ascension limits due to observing time constraints. We perform the additional magnitude cut of $m_{r}<15.7$ to reduce integration times and maximise signal to noise. At our average redshift of $z\sim0.03$, this corresponds to an absolute magnitude of $M_{r}=-20$. Table \ref{ObsTable} lists all galaxies observed for this work. It is important to note that while our target galaxies have been drawn from the very central regions of voids, this does not mean they are in regions free of large-scale structure; tendrils and even filaments of galaxies have been shown to extend deep into  voids. \citet{Alpaslan14} showed that the deeper the magnitude cut, the more structure was revealed in voids, stating as few as 25 per cent of the original void galaxies defined by \citet{Pan12} are actually isolated. 
 

\subsection{Motivation - Are the reddest of massive void galaxies secularly evolving?} 
\label{Motivation}
Despite their similar environmental conditions, void galaxies do not constitute an homogeneous population.
Figure~\ref{SFMS} shows the star forming `main sequence' for all GAMA survey galaxies with $z<0.05$.  Void galaxies from the \citet{Penny15} sample above our mass cut of $\textrm{M}_{\star}>10^{10}\textrm{M}_{\odot}$ and apparent magnitude limit of $m_{r}<15.7$ are also included as orange squares. 
The galaxies observed for this work with available GAMA SFRs are shown as yellow stars, and those without instead with Wide-field Infrared Survey Explorer (WISE) $W3$ band SFRs are shown as purple stars. The WISE SFRs  are H$\alpha$ SFRs derived from the 12$\mu$m luminosity relation of \citet{Cluver14}. For void galaxies above the mass cutoff, both isolated and central void galaxies show a spread in SFR of several dex. The presence of these void galaxies in both the red sequence and blue cloud at present day indicates a spread in quenching times and assembly histories, even for galaxies of similar mass. 

A colour-magnitude diagram of the massive void galaxies reveals just 13 per cent are classified as red-sequence objects using the \citet{Baldry04} line. If we apply our magnitude limit of $m_{r}<15.7$, just nine red sequence and seven blue cloud galaxies remain. We chose to target the reddest galaxies of this sample for our study, as they are the best place to look for the occurrence of mass quenching and constitute an interesting subsample, since in general, void galaxies tend to be bluer than the field. We wish to determine whether these galaxies have received enough external stimulus to evolve into early-type galaxies to match their red colour, or if we are studying a sample of pristine disc galaxies in the process of secular evolution only. While these galaxies have low H$\alpha$ fibre fluxes consistent with optically red galaxies in Figure~\ref{SFMS}, their mid-IR colours suggest a moderate excess of 12$\mu$m emission for at least half of the galaxies targeted. 

\begin{figure}
  
  \centering
\begin{subfigure}{0.4\textwidth}
\includegraphics[width=\textwidth]{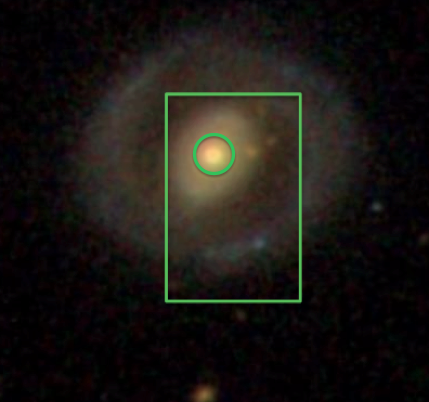}
\caption{An SDSS colour image of the galaxy CGCG 013-075 with SDSS fibre diameter and WiFeS rectangular field of view overlaid in green. While the SDSS fibre catches only the red, passive central bulge region of the galaxy, the WiFeS IFS summed spectra catch the blue spiral arm regions and associated H$\alpha$ emission.}
\label{no10}

\end{subfigure}
\vskip\baselineskip
\begin{subfigure}{0.4\textwidth}
\centering
\includegraphics[width=\textwidth]{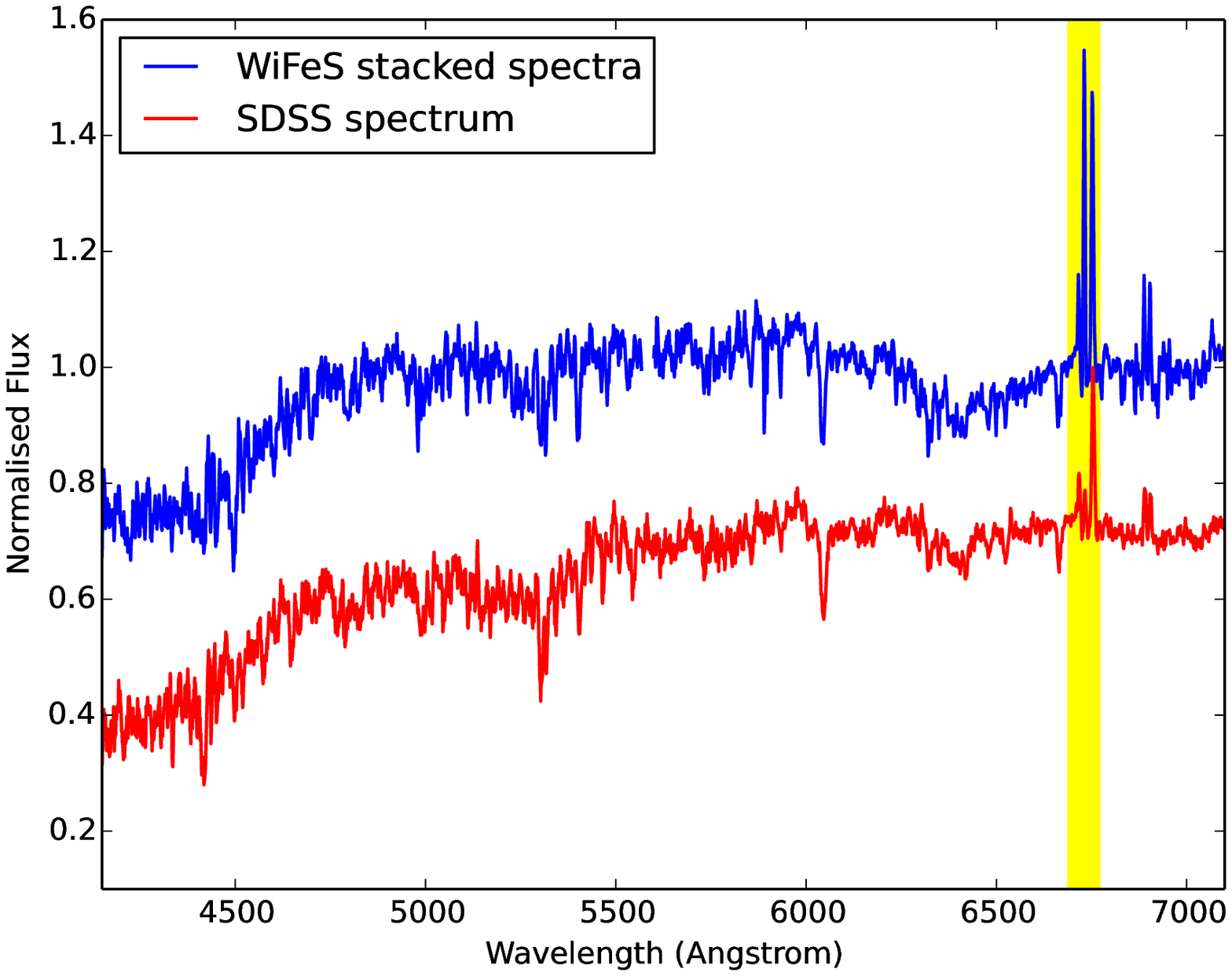}

\caption{The WiFeS IFS summed spectra of CGCG 013-075 with each spaxel velocity-shifted to the centre spaxel (blue) and the SDSS fibre spectrum (red). The H$\alpha$ and [NII] doublet region are highlighted in yellow. The lack of H$\alpha$ in SDSS spectrum is obvious, illustrating how important capturing an entire galaxy is, and how aperture bias may affect SFR estimates for large, nearby galaxies.} 
\label{no9}

\end{subfigure}
\caption{An example of what can be missed with traditional fibre spectroscopy, CGCG 013-075 - a barred spiral galaxy, and its corresponding spectra taken with both a small aperture and an IFS.}
\label{voidgal}
\end{figure}

We wish to examine our massive void galaxy sample using integral field spectroscopy since traditional fibre spectroscopy will often miss regions of star formation in the outskirts of large, nearby galaxies \citep[e.g.,][]{Pracy14}. 
As an example, we present the central void galaxy CGCG 013-075 in Figure~\ref{voidgal}. This galaxy is a face-on, barred spiral with blue, star forming arms as shown in the SDSS colour image (Figure~\ref{no10}). The SDSS fibre diameter is 3$\arcsec$, and the spectrum of this central region of the galaxy is plotted in Figure~\ref{no9} in red, and 950 stacked spaxels from the WiFeS IFS with a field of view of $25\arcsec \times38\arcsec$ and covering the entire galaxy out past an effective radius is plotted in blue. The H$\alpha$ region of the spectra is highlighted in yellow. We see that the H$\alpha$ line is much more pronounced in the IFS spectrum, which includes much of the galaxy, rather than just the central 1.5$\arcsec$, as in the SDSS spectrum. This is a `typical', discy void galaxy, with a red bulge, but star formation occurring in the outer regions. Galaxies such as this may be common in void regions, where interactions may not have been frequent enough to remove star forming discs. Could the void central galaxies in Figure~\ref{SFMS} be mis-plotted on the star forming main sequence due to small fibre size missing star forming regions?
It is for these reasons we seek to spatially examine massive void properties.

 \subsection{How Pristine are these Void Galaxies?}
   We check the likelihood that these galaxies have lived their entire life in voids, and have not been ejected from large scale structure.
   To do this we estimate the maximal distance a galaxy can travel in a Hubble time. If we assume a galactic expansion rate of $\sim$10 Mpc /$H_{t}$, and the exemplar peculiar motion of a galaxy over a Hubble time is $\sim$2-3 Mpc, the furthest a galaxy can travel over a Hubble time is the addition of these two terms, $\sim$12-13 Mpc. \citet{Pan12} state the average void radius is 17$h^{-1}$ Mpc ($\sim$12 Mpc using $H_{0}$=70 $\textrm{km}~\textrm{s}^{-1}$).
   So while it is physically possible that galaxies from large-scale structure may end up in voids, it is not likely, as over time, galaxies are generally gravitationally attracted towards regions of large-scale structure. While it is feasible a low-mass dwarf galaxy could be flung into a void, the chance that one of our high-mass central void galaxies has been cast out of large-scale structure is low, as it would have had to occurred at very early times and move in the correct direction to come to be located in the central regions of a void at current times.
   We may be reasonably confident that all central void galaxies have lived their entire life in voids. While this would point to them having experienced secular evolutionary processes their entire lives,  we stress that this does not preclude interaction with other void galaxies \citep[e.g.,][]{Hirschmann13}.
  
  \section{Observations \& Data Reduction}
  \label{observations}
  Observations were taken over two, six night periods from 2014 March $5^{\textrm{th}}$--$10^{\textrm{th}}$ and 2015 March $23^{\textrm{th}}$--$28^{\textrm{th}}$ using the Wide Field Spectrograph \cite[WiFeS;][] {Dopita07,Dopita10} on the Australian National University's 2.3m telescope at Siding Spring Observatory. 
 WiFeS is a dual beam, image slicing IFS with a field of view of $25\arcsec \times 38\arcsec$, consisting of 25 $\times 1\arcsec$ slitlets of 38$\arcsec$ length. If the data are binned 2$\times$1 in the y direction, the resulting spaxel size is $1\arcsec \times 1\arcsec$. The B3000 ($\sim3500-5800 \textrm{\AA}$) and R3000 ($\sim5300-9000 \textrm{\AA}$) gratings were used along with the RT560 dichroic, resulting in a spectral resolution of $\sigma\simeq100~\textrm{km}~\textrm{s}^{-1}$ for both the red and blue arms. This wavelength range for our low-$z$ target sample includes the H$\alpha$, H$\beta$, H$\delta$ and H$\gamma$ Balmer lines, along with forbidden lines [NII], [SII], [OIII], the 4000~\AA~break and the H \& K absorption line features.
 The 2014 run used WiFeS in `classical' mode, while `nod-and-shuffle' was used in 2015. Both modes spend equal time on target and sky regions, but nod-and-shuffle mode takes shorter exposures, moving back and forth between object and sky regions more rapidly, providing a more real-time sky spectrum, and facilitating better sky subtraction.
  Weather conditions were a mixture of clear and patchy cloud, with an average seeing of $2\arcsec$ in non-photometric conditions. Each field was observed a minimum of three times, to allow removal of cosmic rays via median recombination. Table \ref{ObsTable} lists all void galaxies observed.
  
  The data were reduced using the dedicated WiFeS data reduction pipeline \citep[\textsc{pywifes};][]{Childress14}. The \textsc{python} routine includes bias subtraction, flatfielding, wavelength calibration, sky subtraction, cosmic ray removal and flux calibration using spectrophotometric standard stars. The final product is two data cubes, one red and one blue, for each galaxy. 
  
\begin{sidewaystable}
\caption{Observations and SFR comparisons for the nine central void galaxies in our sample.}
\label{ObsTable}
\centering
\begin{tabular}{l c c c c c c c c c c}
\hline
\textbf{Galaxy} & \textbf{redshift}  &  \textbf{Obs date}  & \textbf{exp time} & \textbf{$m_{r}^{1}$}   &\textbf{$u-r^{2}$} & \textbf{Stellar Mass}$^{3}$   & \textbf{Parent}& \textbf{H$\alpha$ SFR}$^{4}$  & \textbf{WISE SFR}$^{5}$                   & \textbf{GAMA SFR}$^{6}$   \\
                          &   &                &           \textbf{(s)}                   &        &              &  \textbf{($\times 10^{10} \textrm{M}_{\odot}$)}  & \textbf{survey}              &\textbf{($\textrm{M}_{\odot}~\textrm{yr}^{-1}$)} &\textbf{($\textrm{M}_{\odot}~\textrm{yr}^{-1}$)} & \textbf{($\textrm{M}_{\odot}~\textrm{yr}^{-1}$)}   \\
\hline
  IC 0566               & 0.03660 &  Mar 2015  & 4800 &  14.15 &  2.55 & 10.71  & SDSS & 0.03 & 0.22   &      \\ 
  CGCG 010-071  & 0.03362 & Mar 2014  &  3600   &13.92  & 2.87 & 10.92   & SDSS & 0.24 & 1.34     &      \\ 
  GAMA 214363 & 0.04213 & Mar 2015   &4500 & 14.50 & 2.77 &10.96 & GAMA & 0.06 & 0.14   & --    \\ 
  GAMA 534760  & 0.05259 &  Mar 2015  & 6000 & 15.72 & 2.82 & 10.74 & GAMA & 0.09 & 0.02    &  0.55    \\ 
  CGCG 013-075  & 0.02558 &  Mar 2015  & 5633 & 14.05 & 2.66 & 10.79 & SDSS & 0.45 & 0.72   &       \\  
  CGCG 019-064  & 0.01393 &  Mar 2015  & 4200  & 14.04 & 2.35 & 10.12 & GAMA & 0.05 &0.19    & 0.20     \\ 
  IC 1059              & 0.02646 &   Mar 2015 & 4080  & 13.69  & 2.74 & 10.83 & GAMA & 0.48 & 0.31    & 0.31      \\ 
  IC 0653               & 0.01847 &   Mar 2014  &  3600  & 12.90  & 2.71 & 10.62 & SDSS & 0.26 & 2.33   & --     \\ 
 CGCG 017-063  & 0.02864 &   Mar 2014  &  2700  & 13.42  & 2.67 & 10.75 & SDSS & 0.70 & 3.29  & --      \\ 
\hline
\multicolumn{11}{l}{$^{1}$SDSS r-band model magnitudes.}\\
\multicolumn{11}{l}{$^{2}$SDSS model magnitudes. All galaxies have $u-r$ colour such that they would be located on the red sequence as defined by \citet{Baldry04}.}\\
\multicolumn{11}{l}{$^{3}$From \citet{Taylor11}.}\\
\multicolumn{11}{l}{$^{4}$Total integrated SFRs from this work.}\\
\multicolumn{11}{l}{$^{5}$12$\mu$m SFR relation from \citet{Cluver14} adjusted to a \citet{Salpeter55} IMF using the conversion of \citet{Gunawardhana13}.}\\
\multicolumn{11}{l}{$^{6}$ From \citet{Hopkins13}, adjusted to a \citet{Salpeter55} IMF using the conversion of \citet{Gunawardhana13}.}\\

\end{tabular}

\end{sidewaystable}

  \section{Stellar and Gas Kinematics}
  \label{stellar}
\begin{figure}
 \centerline{ 
\psfig{file=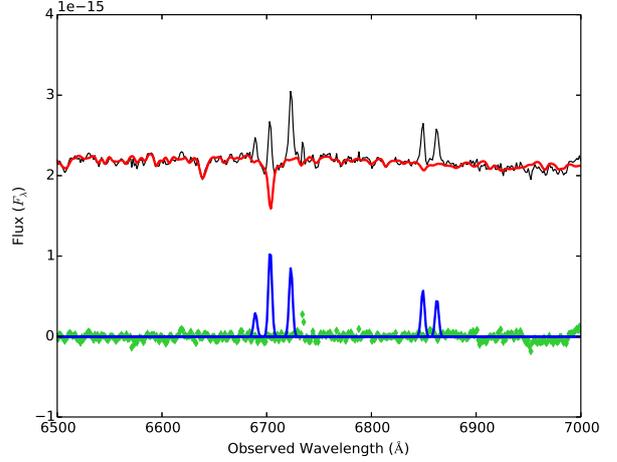,angle=0,width=3.5in}
}
\caption{An example of a pPXF fit for the central spaxel of the void galaxy CGCG 019-064 around the H$\alpha$ region (second emission line from the left). The original galaxy spectrum is in black, the best fit pPXF stellar continuum fit in red, single gaussian-fitted emission lines in blue, and residuals in green. Note the extra emission flux added to the H$\alpha$ line after correcting for flux lost due to stellar Balmer absorption.} 
\label{pPXFexample}
\end{figure}

 The stellar and gas kinematics of each galaxy (velocity and line-of-sight velocity dispersion, $\sigma$), were analysed using the penalized pixel fitting (pPXF) method of \citet{Cappellari04}. To improve the signal to noise (S:N) of the outer regions of the galaxy, the data cubes were first Voronoi binned to obtain a S:N of 15 per bin for the red arm and 5 per bin for the blue arm using the \textsc{python} script of \citet{Cappellari03}. The difference in S:N cuts is because the WiFeS blue arm camera is less sensitive than the red. 
\citet{Cappellari04}'s Voronoi binning technique accretes spaxels (beginning at the brightest spaxel of a galaxy) into a bin until the target S:N is reached with constraints on bin shape.  While the central spaxels generally provided high enough S:N, most outer regions of low galaxy surface brightness needed significant binning. As outer regions comprised bins of several spaxels which were not velocity shifted before analysis we expect some smearing of emission lines affecting kinematic analysis. These bins also didn't usually reach the target S:N and were therefore not used in our emission line analyses however. 
  The stellar continuum was modelled using a subset of the Indo-U.S. Library of Coud\'e Feed Stellar Spectra templates \citep{Valdes04} covering a range of stellar ages and metallicities. These templates were chosen for their long wavelength coverage (3460$\textrm{\AA}$ \-- 9464$\textrm{\AA}$) and high spectral resolution (FWHM = 1.35$\textrm{\AA}$). Gas emission lines were fitted simultaneously to avoid the need for masking by modifying the example \texttt{ppxf\_population\_gas\_example\_sdss.py} provided in the pPXF package. Emission lines were fitted using a single Gaussian profile. Figure~\ref{pPXFexample} shows an example pPXF fit to the central spaxel of the void galaxy CGCG 019-064 around the H$\alpha$ region. 

An error approximation for the stellar kinematics was completed by performing 100 Monte Carlo realisations of the best fit output by pPXF for each Voronoi bin of every galaxy. To do this, random noise is added to the best fit, then the pPXF fit performed 100 times. The standard deviation of the weighted mean of all the fits with the random noise added is our uncertainty in our velocity, velocity dispersion and emission line values.

\section{Fluxes, equivalent widths and star formation rates}
\label{SFRsection}
Emission line fluxes are calculated by numerically integrating the area underneath the pPXF emission line fits. H$\alpha$ emission line equivalent widths (EWs) are calculated from these flux values and the continuum levels $\sim 30 \textrm{\AA}$ either side of the H$\alpha$ line.\\
As we have modelled and subtracted the stellar continuum using pPXF, we do not need to include any stellar absorption corrections, though we must account for obscuration by dust in our derived star formation rates. 
We calculate the H$\alpha$-derived star formation rates (SFR) of our galaxies by using the SFR relation employed by \citet{Richards15}:
\begin{equation}
\label{eq1}
\textrm{SFR}(\textrm{M}_{\odot}~\textrm{yr}^{-1}) = \frac{\textrm{F}_{\textrm{H}_{\alpha}} \times 4 \pi  \textrm{D}_{L}^{2}}{1.27 \times 10^{34}\textrm{W}} \times \left( \frac{\textrm{BD}}{2.86} \right)^{2.36},
\end{equation}
where $\textrm{F}_{\textrm{H}_{\alpha}}$ is the integrated H$\alpha$ flux in $\textrm{W}~\textrm{m}^{-2}$, $\textrm{D}_{L}$ is the luminosity distance in m, and $1.27 \times 10^{34}\textrm{W}$ is the conversion factor to convert to solar masses per year, as given by \citet{Kennicutt98} assuming a \citet{Salpeter55} IMF and solar metallically. BD is the Balmer decrement of the stellar absorption-corrected ratio of H$\alpha$ to H$\beta$ emission line fluxes assuming a Case-B recombination value of 2.86 \citep{Calzetti01} with a reddening slope of 2.36 \citep{Cardelli89} and the dust as a foreground screen averaged over the galaxy. Balmer decrement values varied from spaxel to spaxel throughout each galaxy, and where either H$\alpha$ or H$\beta$ measurements weren't available for a given spaxel, we take the average Balmer decrement of the neighbouring spaxels. As an example, for the galaxy CGCG 013-075, Balmer decrement values span the range from $\sim$2.9 on the outskirts to 6.5 in the central regions.

The SFR is calculated for each Voronoi bin above the target S:N, then added together to get the total SFR for a galaxy.
Table~\ref{ObsTable} compares the SFRs calculated in this work with optical (GAMA) and mid-IR (WISE) integrated SFRs from the literature. 
The WISE SFRs are calculated using the relation of \citet{Cluver14}, which is derived using the GAMA SFR relation of \citet{Wijesinghe11}, assuming the IMF definition of \citet{Baldry03}. Stellar emission has been subtracted from the $W3$ band using the $W2$ band, leaving just emission from hot dust associated with star formation.
\citet{Gunawardhana13} find the difference between the GAMA IMF and a \citet{Salpeter55} IMF results in a SFR approximately a factor of 2.4 lower.
We therefore multiply the \citet{Cluver14} SFRs by a factor of 2.4, for a comparison to the values presented in this work using the \citet{Richards15} relation.

\section{Results \& Discussion}
\label{results}
\subsection{Secularly Evolving Void Galaxies}
\begin{figure*}
\centering
\begin{subfigure}{0.3\textwidth}
\includegraphics[width=\textwidth]{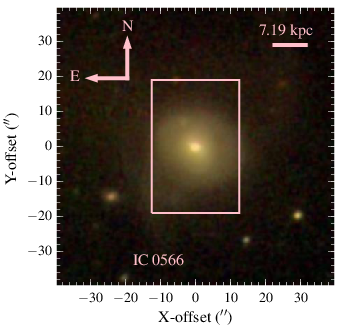}
\caption{IC 0566 }
\end{subfigure}
\hfill
\begin{subfigure}{0.3\textwidth}
\centering
\includegraphics[width=\textwidth]{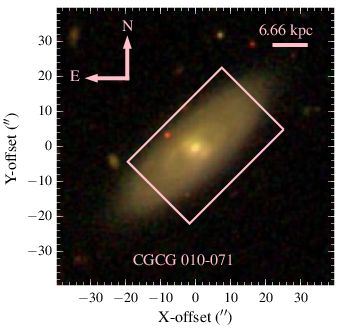}
\caption{CGCG 010-071}
\end{subfigure}
\hfill
\begin{subfigure}{0.3\textwidth}
\centering
\includegraphics[width=\textwidth]{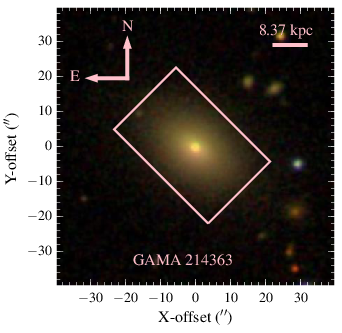}
\caption{GAMA 214363}
\end{subfigure}
\vskip\baselineskip
\begin{subfigure}{0.3\textwidth}
\centering
\includegraphics[width=\textwidth]{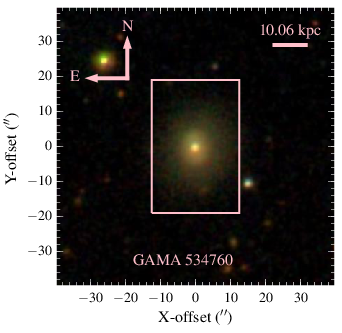}
\caption{GAMA 534760}
\label{no3}
\end{subfigure}
\hfill
\begin{subfigure}{0.3\textwidth}
\centering
\includegraphics[width=\textwidth]{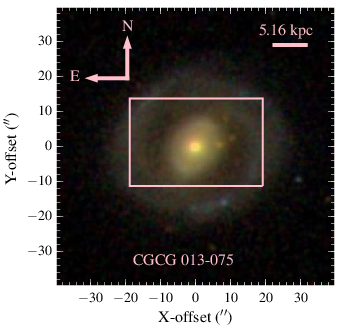}
\caption{CGCG 013-075}
\end{subfigure}
\hfill\begin{subfigure}{0.3\textwidth}
\centering
\includegraphics[width=\textwidth]{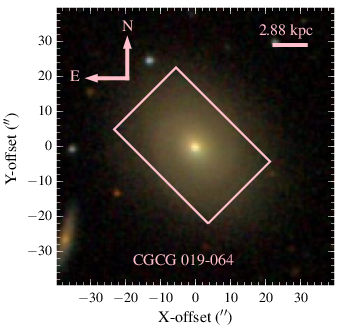}
\caption{CGCG 019-064}
\end{subfigure}
\vskip\baselineskip
\begin{subfigure}{0.3\textwidth}
\centering
\includegraphics[width=\textwidth]{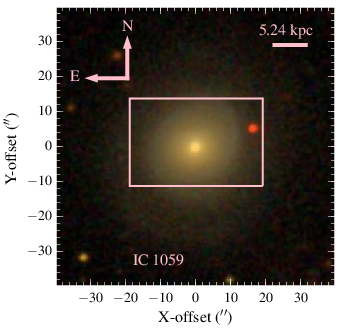}
\caption{IC 1059}
\end{subfigure}
\hfill
\begin{subfigure}{0.3\textwidth}
\centering
\includegraphics[width=\textwidth]{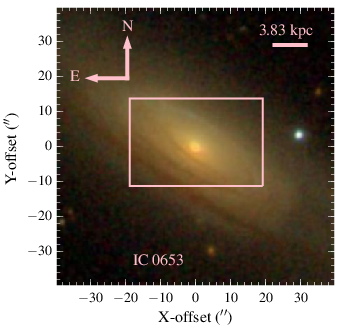}
\caption{IC 0653}
\end{subfigure}
\hfill\begin{subfigure}{0.3\textwidth}
\centering
\includegraphics[width=\textwidth]{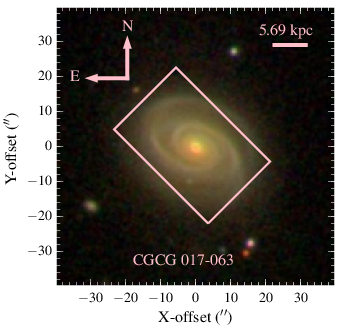}
\caption{CGCG 017-063}
\end{subfigure}
\hfill
\caption{SDSS colour images of the nine void galaxies in the sample with the WiFeS field of view of $25\arcsec\times38\arcsec$ overlaid in pink. All galaxies appear optically red, with the exception of CGCG 013-075, which possesses blue spiral arms. IC 0566 contains a total tail in the South-East corner, likely the result of a recent merger.}
\label{SDSS_gals}
\end{figure*}

  Firstly, we morphologically classify our galaxies, using the SDSS colour images shown in Figure~\ref{SDSS_gals}. We see that eight of the nine galaxies observed have obvious discs. Of these, we visually classify four galaxies as spiral, one of which has blue arms, the other three are red. Four galaxies have lenticular morphology, and the final galaxy is disturbed. 
    
\begin{figure*}
 \centerline{ 
\epsfig{file=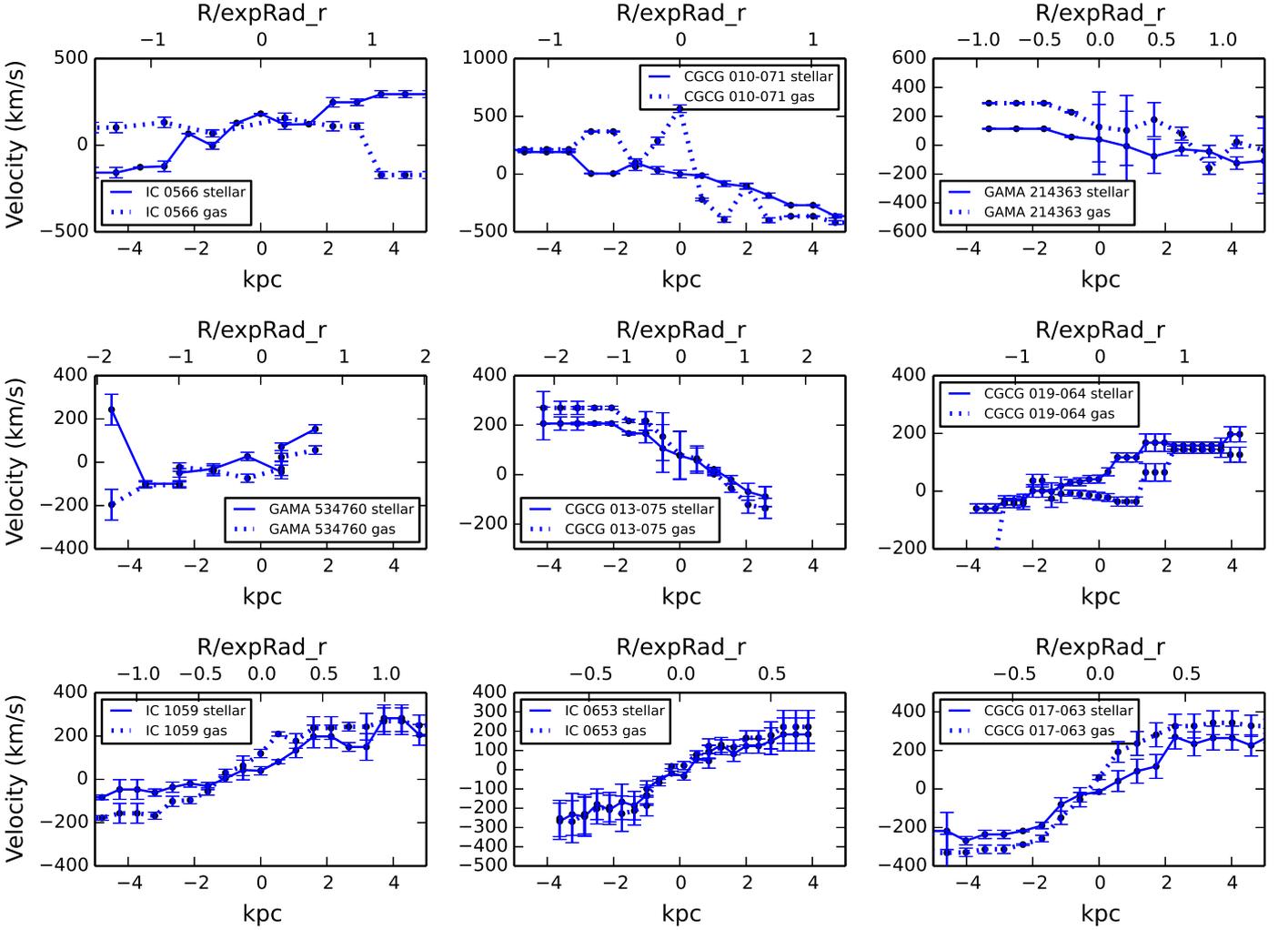,angle=0,width=7.5in}
}
\caption{Stellar (solid line) and gas (dashed line) rotation curves as measured along each galaxy's kinematic major axis as a function of both distance from galaxy centre (kpc) and fraction of the SDSS exponential fit radius, expRad\_r, corrected for galaxy inclination. Error bars denote the MC error for the pPXF fit to the Voronoi bin. Kinematics derived from Voronoi bins that didn't reach the target S:N (those on the edges) are not included. The gas and stellar kinematics, both derived using pPXF, match up well in the central regions (where there is high S:N) for all but two galaxies - IC 0566 and CGCG 010-071, which we expect merger activity and a galactic bar disrupting  gas flow to be the cause of the obviously disturbed gas morphologies.}
\label{rot_curve}
\end{figure*}

Our morphological classifications are confirmed by Figure~\ref{rot_curve}, where we present the stellar (solid line) and gas (dashed line) velocity curves for all void galaxies in our sample extracted along the kinematic major axis of each galaxy. The stellar kinematics show all nine galaxies are rotating regularly with a maximum velocity of several hundred $\textrm{km}~\textrm{s}^{-1}$. The rapid rotation implies these galaxies are (or recently have been) discs. The discy nature of the galaxies in our sample coupled with high stellar mass implies isolation, and a secular evolutionary history.  

\begin{figure*}
 \centerline{ 
\epsfig{file=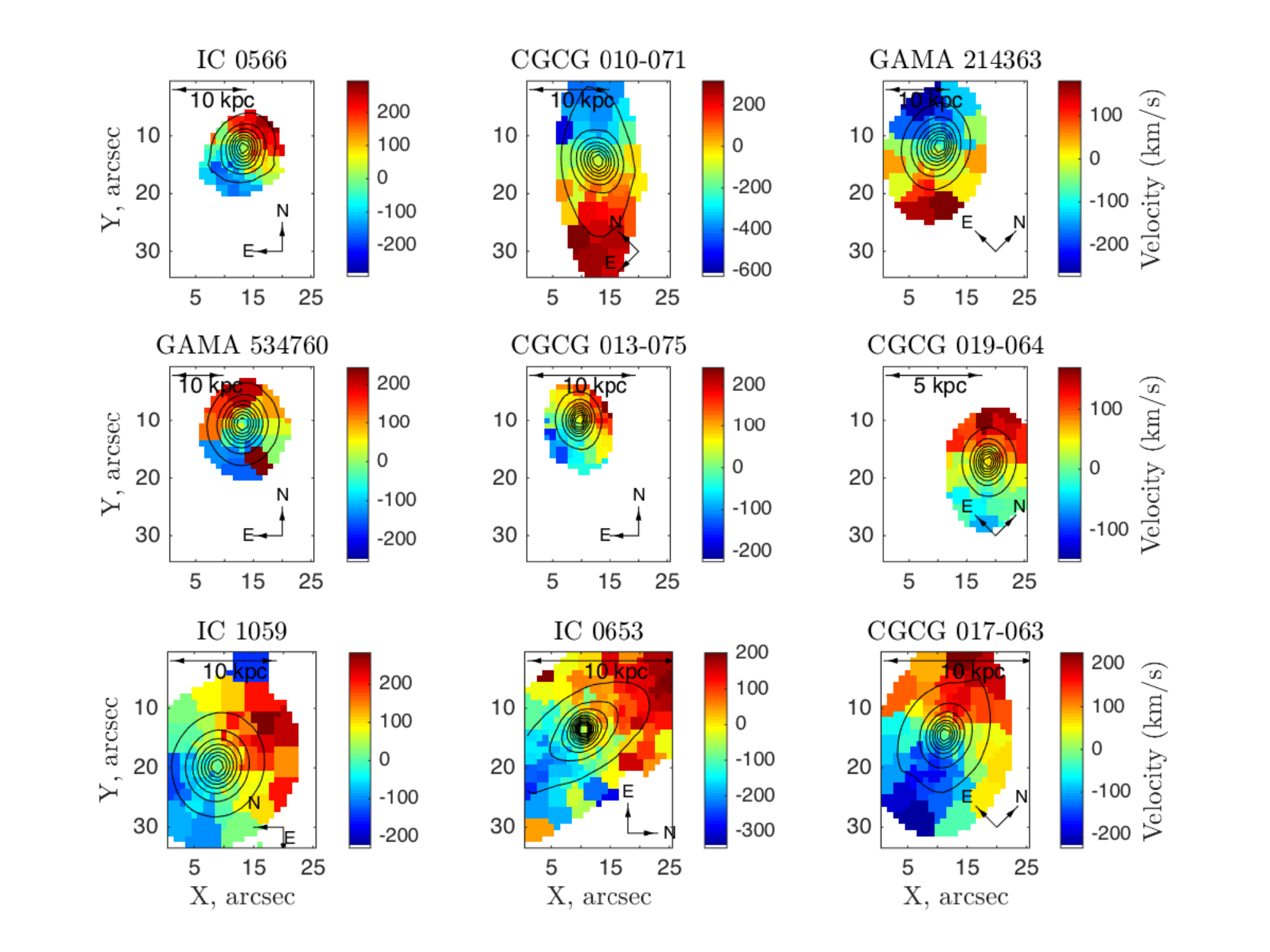,angle=0,width=6.5in}
}
\caption{Stellar velocity maps of the nine void galaxies in our sample derived from the pPXF fit to each Voronoi bin above the S:N threshold. Total integrated flux contours are overlaid in black. All galaxies are rotating regularly, leading them to be classified as discs.}
\label{stell_maps}
\end{figure*}

\begin{figure*}
 \centerline{ 
\epsfig{file=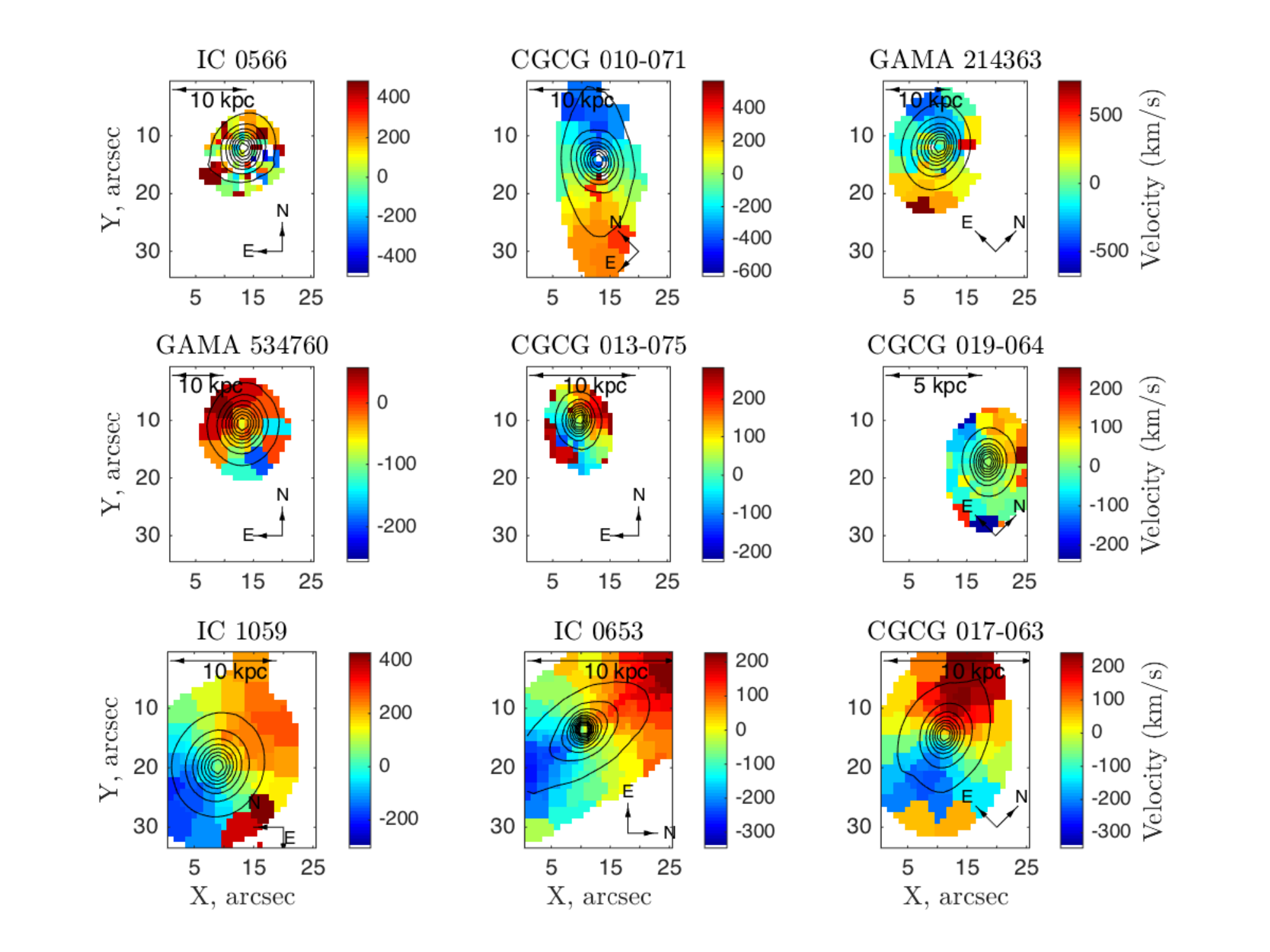,angle=0,width=6.5in}
}
\caption{Gas velocity maps for the nine void galaxies in the sample derived from the pPXF fit to the nebular emission lines above the S:N threshold. If nebular emission lines are not detected by the pPXF fit for a region, the gas velocity is not measured, as is the case for some central regions of IC 0566 and CGCG 010-071. For the majority of galaxies the gas and stellar velocities match well, though the central region of CGCG 010-071 is disturbed. The gas of IC 0566 does not match the stellar velocity map, leading us to conclude this galaxy has experienced an event to disrupt the gas, such as a merger. }
\label{gas_maps}
\end{figure*}

In Figures~\ref{stell_maps} and \ref{gas_maps} we present maps of stellar and gas kinematics respectively. The constant rotation of the stellar kinematics further confirm the discy nature of these galaxies. The gas kinematics matches the stellar for most of the galaxies in our sample, leading us to conclude they have had no recent interactions, and confirming their secularly evolving nature. For the galaxy CGCG 010-071, the central regions of the gas and stelar kinematic maps show some discrepancy, leading to central velocity differences between gas and stellar components along the kinematic major axis of several hundred $\textrm{km}~\textrm{s}^{-1}$. \citet{Nair10} classify this galaxy as a barred spiral, and we suggest the gas kinematics are being disrupted in the central regions by the presence of a galactic bar \citep[e.g.,][]{Masters11}.


  From Figure~\ref{SDSS_gals} we see that IC 0566 has an obvious extension in the South West corner, likely the result of a recent merger. Mergers can disturb gas kinematics whilst still maintaining regular rotation of stellar kinematics. Indeed, from Figures~\ref{stell_maps} and \ref{gas_maps}, we see that while the stellar component is rotating regularly, the gas kinematics are highly variable from spaxel to spaxel. As this galaxy has very little nebular emission detected from the pPXF fit in its central regions, the gas velocity can only be determined for the spaxels which have reliable emission line information. Despite this, those with gas velocity information are significantly variable enough to suggest the gas in this galaxy is very disturbed. In combination with the visual asymmetry of this galaxy, we conclude it is undergoing or has recently undergone a merger.  
    We investigate the literature to determine the likely merger scenario that may have occurred.\\
  
  \citet{Hirschmann13} performed a detailed statistical analysis of isolated model galaxies extracted from the Millennium simulation \citep{Springel05}. They found that $\sim45 \%$ of isolated galaxies at $z$=0 have experienced at least one merger event in their lifetimes, most of which were minor ($< 1:4 $ mass ratio). A small fraction of isolated galaxies in their sample were bulge-dominated, whose bulges had been built mainly from minor mergers. 
\citet{Penny15} also used the Millennium simulation to find the time since last merger for isolated, massive void galaxies. They find that the major merger histories of massive void galaxies differ significantly from non-void galaxies. The major merger rates are typically much lower than non-void galaxies and almost negligible at current times. One per cent of void galaxies with $\textrm{M}_{\star}>5\times10^{10}\textrm{M}_{\odot}$ have undergone a major merger in the last 5 Gyr (compared to 21 per cent of the total galaxy population in this mass range), demonstrating that the dominant morphology of massive void galaxies remains discy. A larger fraction of major mergers is effectively ruled out by $\Lambda$CDM cosmology as the cause of the gas disruption in our galaxies, so we assume that minor mergers and gas accretion are the cause. We now investigate the chance an isolated void galaxy will experience an interaction at current times.

Observationally, examples exist of interacting systems in void regions. In fact, the HI discs of galaxies in voids are just as likely to be disturbed as they are undisturbed. The void galaxy survey \citep{Kreckel12} imaged 60 void galaxies in HI to find that $\sim$50 per cent show strongly disturbed gas morphology or kinematics, $\sim$8 per cent of which are clearly interacting with companions. 
 \citet{Beygu13} study an interacting system of three galaxies embedded in a common HI envelope. A tidal tail is seen on one of the galaxies, attributed to a recent minor merger. While this system is likely very unusual, it does contribute along with the interacting galaxies in \citet{Kreckel12} to the anecdotal evidence that minor mergers in voids are still occurring at current times. 
 
 While we did not select our sample based on disturbed morphology, our results confirm those of \citet{Kreckel12}, that interacting void galaxies are not uncommon, and the dominant method of interaction is minor mergers. We investigate current day SFRs in Section~\ref{SF_gal} to determine whether these mergers were likely to have been wet or dry.
 
 \subsection{H$\alpha$ Maps and Star Formation in Massive, Central Void Galaxies}
 \label{SF_gal}
 
 \begin{figure*}
 \centerline{ 
\epsfig{file=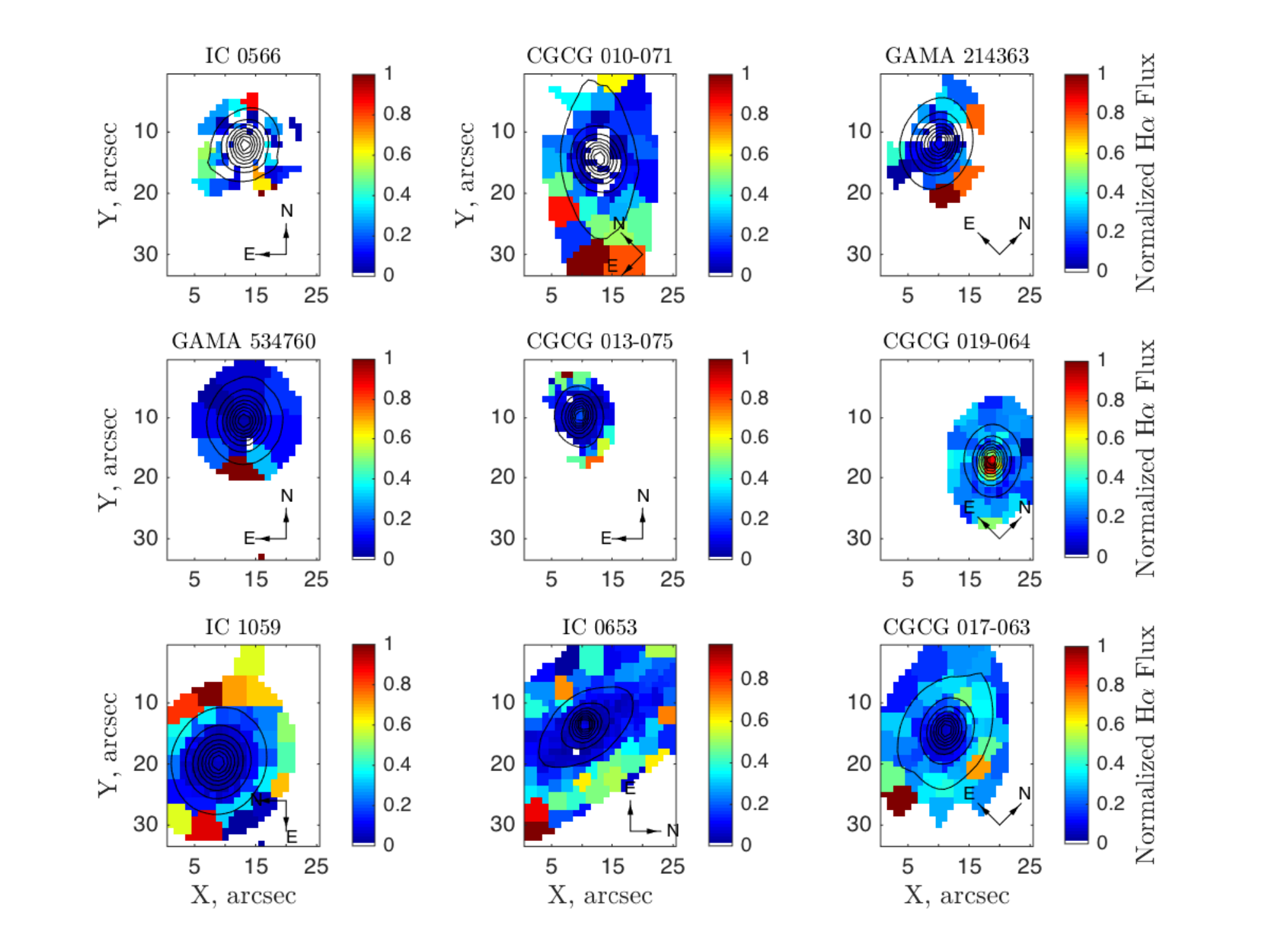,angle=0,width=7.5in}
}
\caption{Normalised H$\alpha$ flux maps for all central regions of the galaxies in our sample. White regions correspond to where the pPXF fit didn't pick up any H$\alpha$ emission. The black contours overlaid are total integrated flux. The majority of our sample possess discs of younger stellar population with old stars in their nuclear region. The exception is CGCG 019-064, which has the majority of its H$\alpha$ emission in its core. This is due either to star formation or AGN emission.}
\label{Ha_maps}
\end{figure*}
In Figure~\ref{Ha_maps} we present the normalised H$\alpha$ emission line maps for the nine void galaxies in our sample. For all but one galaxy (CGCG 019-064) we see a common scenario emerging of a low relative amount of H$\alpha$ emission in the cores of these galaxies, and a younger stellar population in the disc regions of the galaxies. This H$\alpha$ distribution is what we would expect for a disc galaxy population, and confirms the notion that these galaxies are secularly evolving. For CGCG 019-064, we investigate whether the central concentration of H$\alpha$ is either nuclear star formation or AGN activity in Section~\ref{AGN}.

 We calculate SFRs from these H$\alpha$ emission line maps using Equation~\ref{eq1}, and in Table~\ref{ObsTable} we present the total integrated SFR of all galaxies in our sample. All galaxies possess SFRs $< 1~\textrm{M}_{\odot}~\textrm{yr}^{-1}$. 
 These SFRs are surprisingly low, given their predominantly discy morphology. We expect that our total integrated SFRs underestimate the true SFR of the galaxy for two reasons. A minor flux calibration issue was discovered when comparing the WiFeS spectra to their Sloan r-band flux at the r-band effective wavelength. 
 This flux discrepancy is likely due to non-photometric conditions whilst observing. As a result, we find total continuum (and hence H$\alpha$) flux to be underestimated by $\sim20\%$, which propagates to an underestimation of SFR of $\sim20\%$.
Additionally, the Voronoi bins that do not reach the target S:N on the outskirts of the WiFeS field of view were not counted towards the total SFR value. This resulted in a smaller area of the galaxy being covered than anticipated, potentially missing some star forming areas of the disk. As Figure~\ref{Ha_maps} shows, this is where the younger stellar population of the majority of these galaxies lie.
 
 To assess the effect this has on the total SFRs, we compare our integrated SFRs to that of the photometrically-derived SFRs using WISE 12$\mu$m band colour from the relation of \citet{Cluver14}, and GAMA fibre SFRs.
 In general, we find the fibre-derived SFRs are systematically lower than the total integrated SFRs from this work. We expect this is because the aperture corrections applied assume the galaxy outskirts are as red and passive as the nuclear region, as illustrated in Figure~\ref{voidgal}.
In contrast, the WISE SFRs are systematically higher than those calculated in this work. 

SFR discrepancies are most noticeable in the galaxies with large angular size. Reassuringly, our smallest galaxy, GAMA 534760, which fits comfortably in the WiFeS field of view, has a SFR that compares well to the WISE SFRs. Perhaps not surprisingly, since these galaxies are optically red, we do not see any evidence of substantial ongoing levels of star formation.
  
 \subsection{AGN activity}
 \label{AGN}
\begin{figure*}
\centering
\begin{subfigure}{0.3\textwidth}
\includegraphics[width=\textwidth]{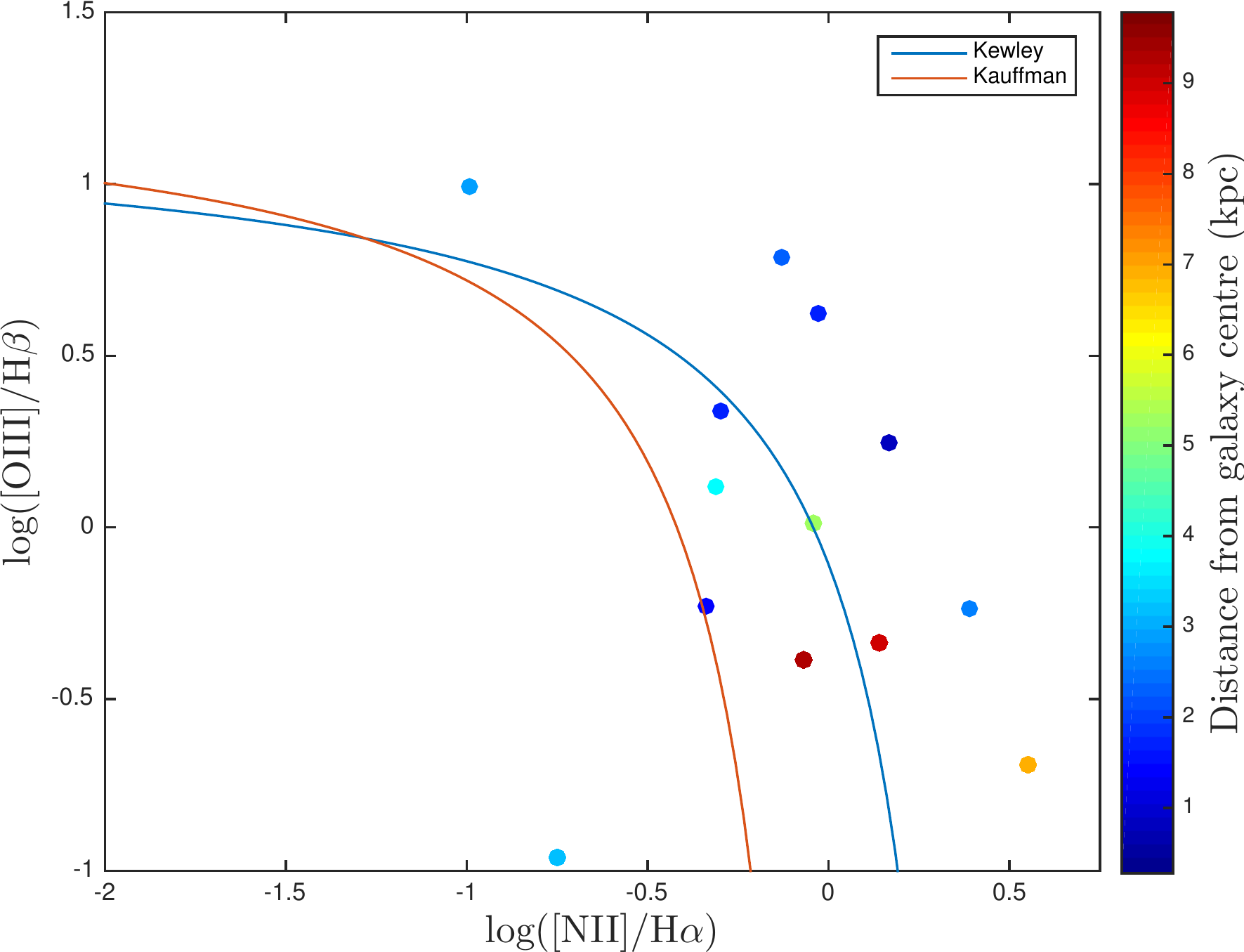}
\caption{IC 0566 }
\end{subfigure}
\hfill
\begin{subfigure}{0.3\textwidth}
\centering
\includegraphics[width=\textwidth]{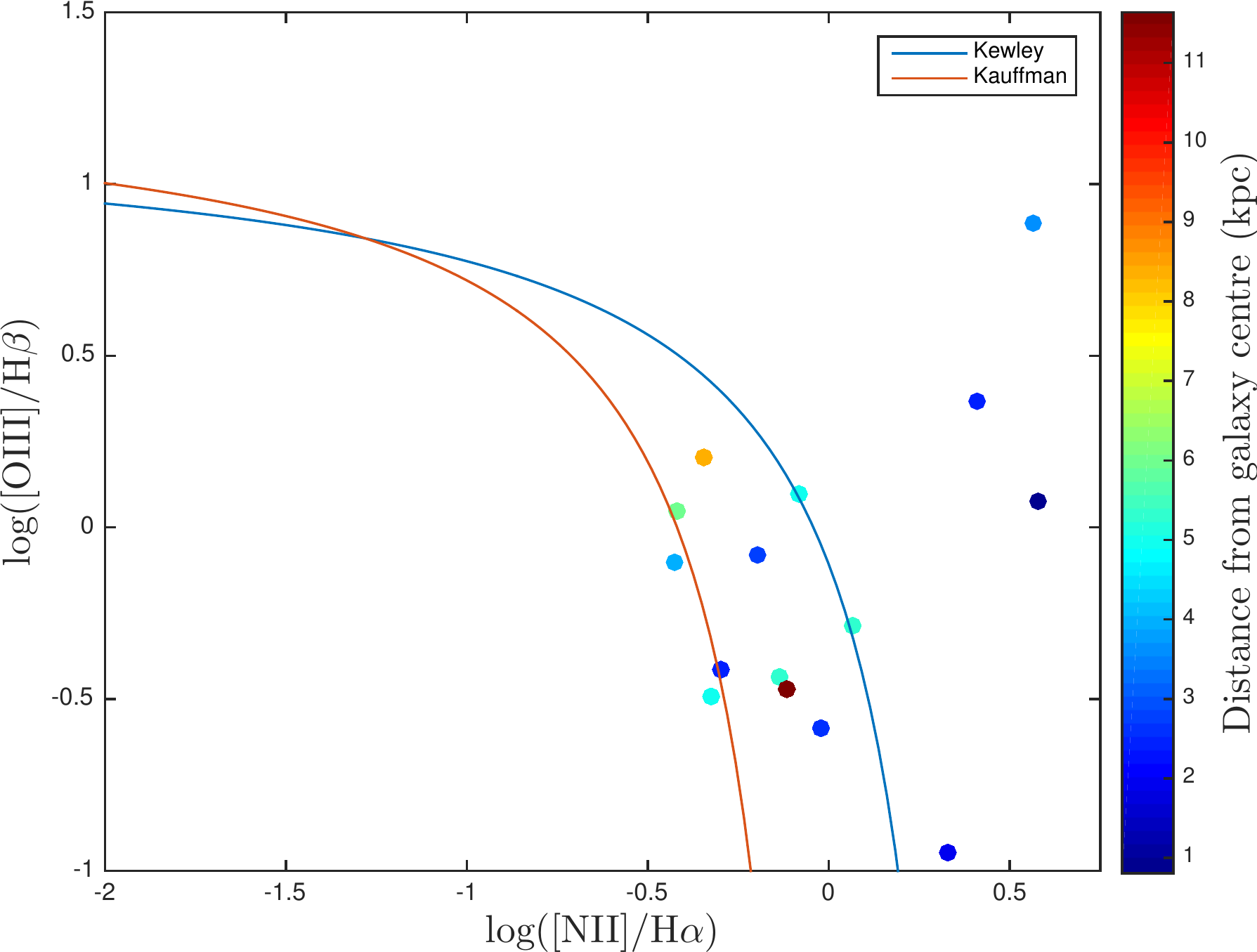}
\caption{CGCG 010-071}
\end{subfigure}
\hfill
\begin{subfigure}{0.3\textwidth}
\centering
\includegraphics[width=\textwidth]{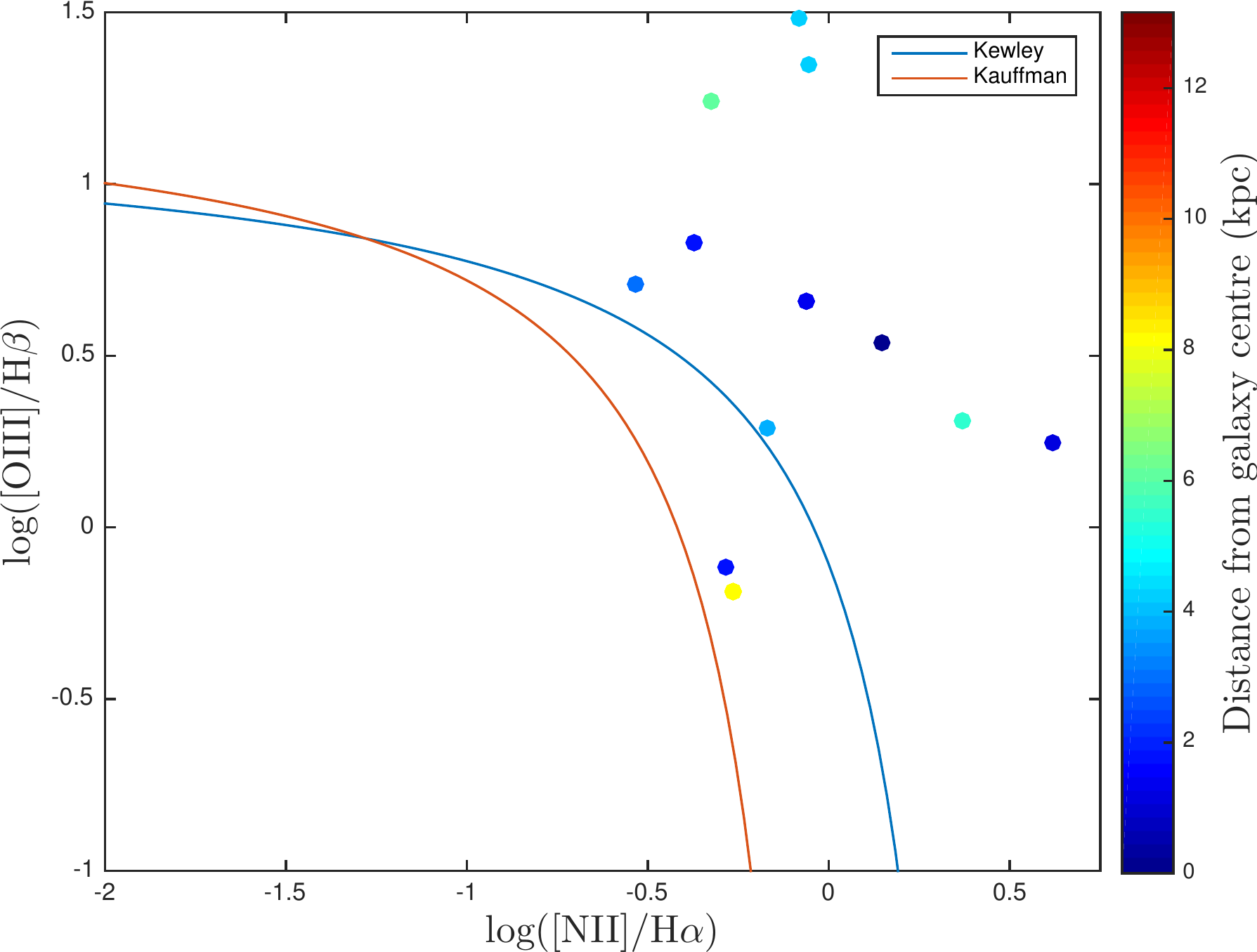}
\caption{GAMA 214363}
\end{subfigure}
\vskip\baselineskip
\begin{subfigure}{0.3\textwidth}
\centering
\includegraphics[width=\textwidth]{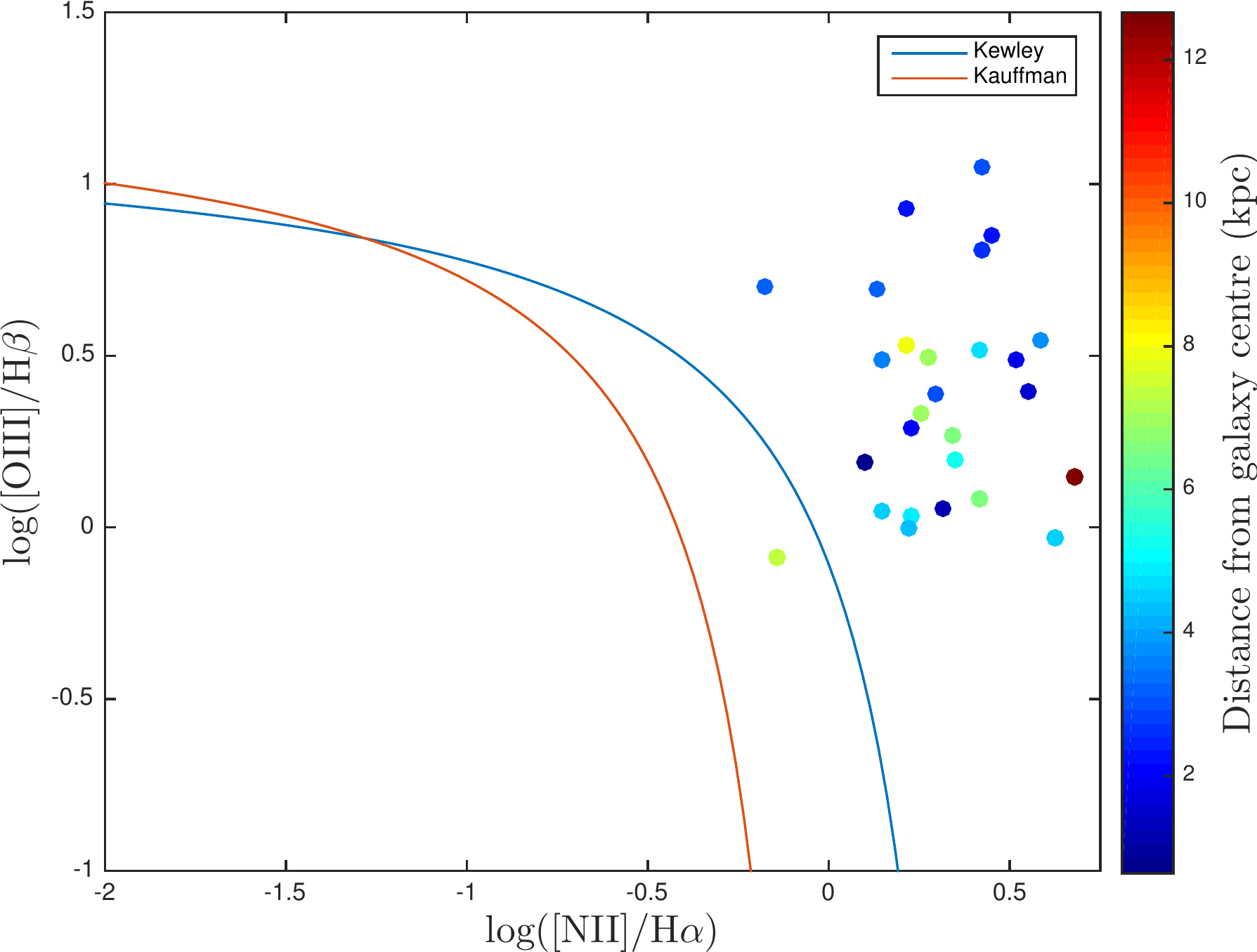}
\caption{GAMA 534760}
\label{no3}
\end{subfigure}
\hfill
\begin{subfigure}{0.3\textwidth}
\centering
\includegraphics[width=\textwidth]{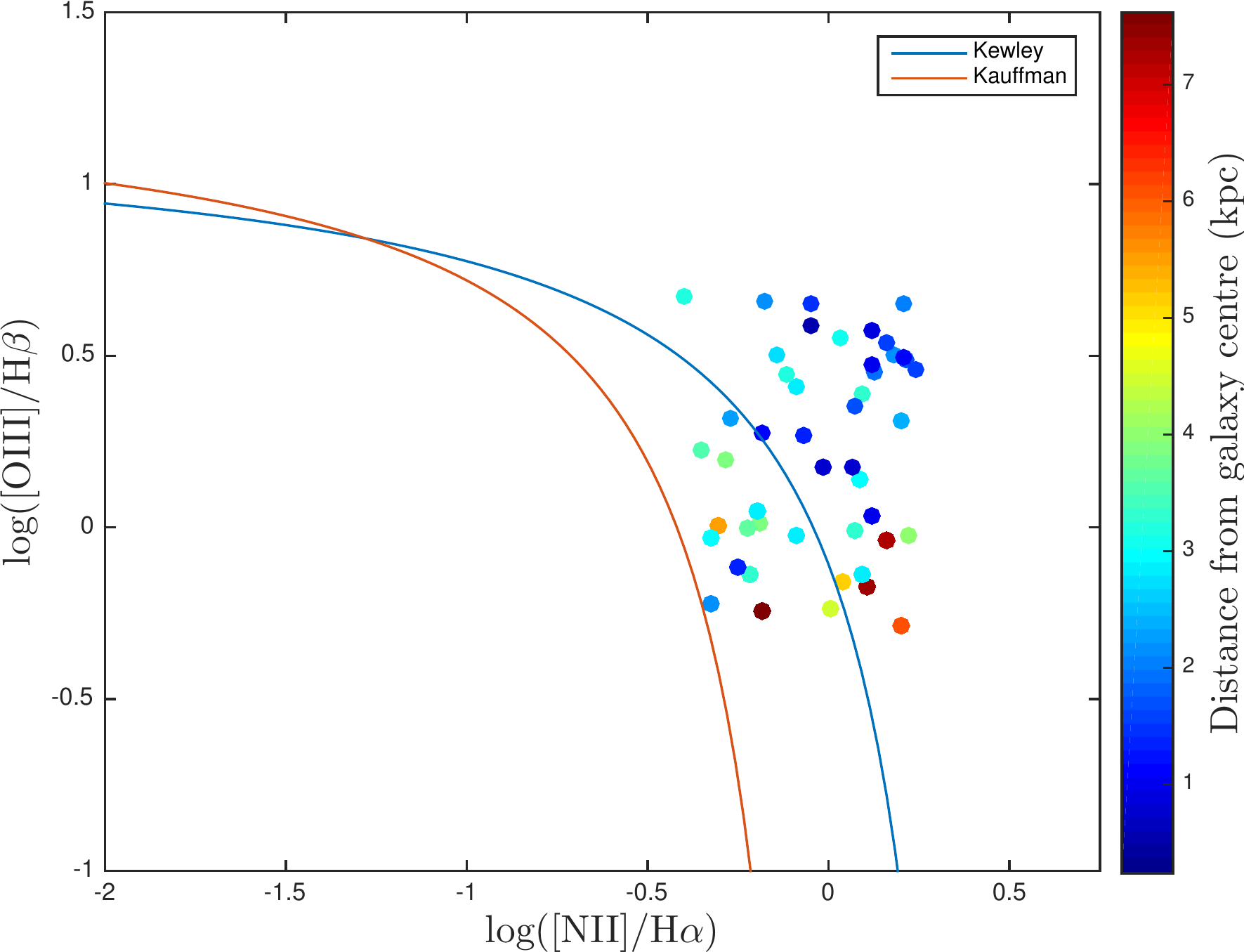}
\caption{CGCG 013-075}
\end{subfigure}
\hfill\begin{subfigure}{0.3\textwidth}
\centering
\includegraphics[width=\textwidth]{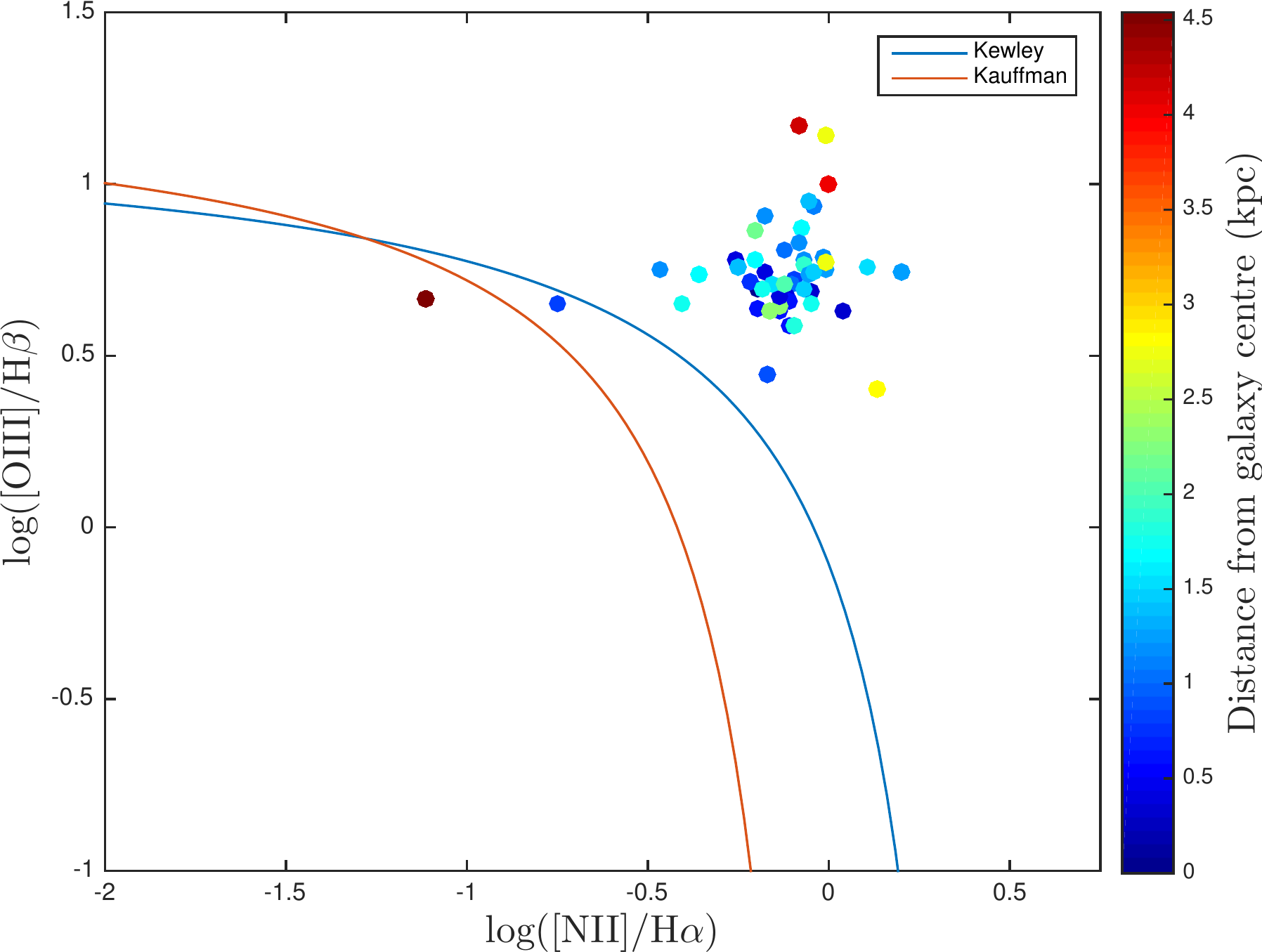}
\caption{CGCG 019-064}
\end{subfigure}
\vskip\baselineskip
\begin{subfigure}{0.3\textwidth}
\centering
\includegraphics[width=\textwidth]{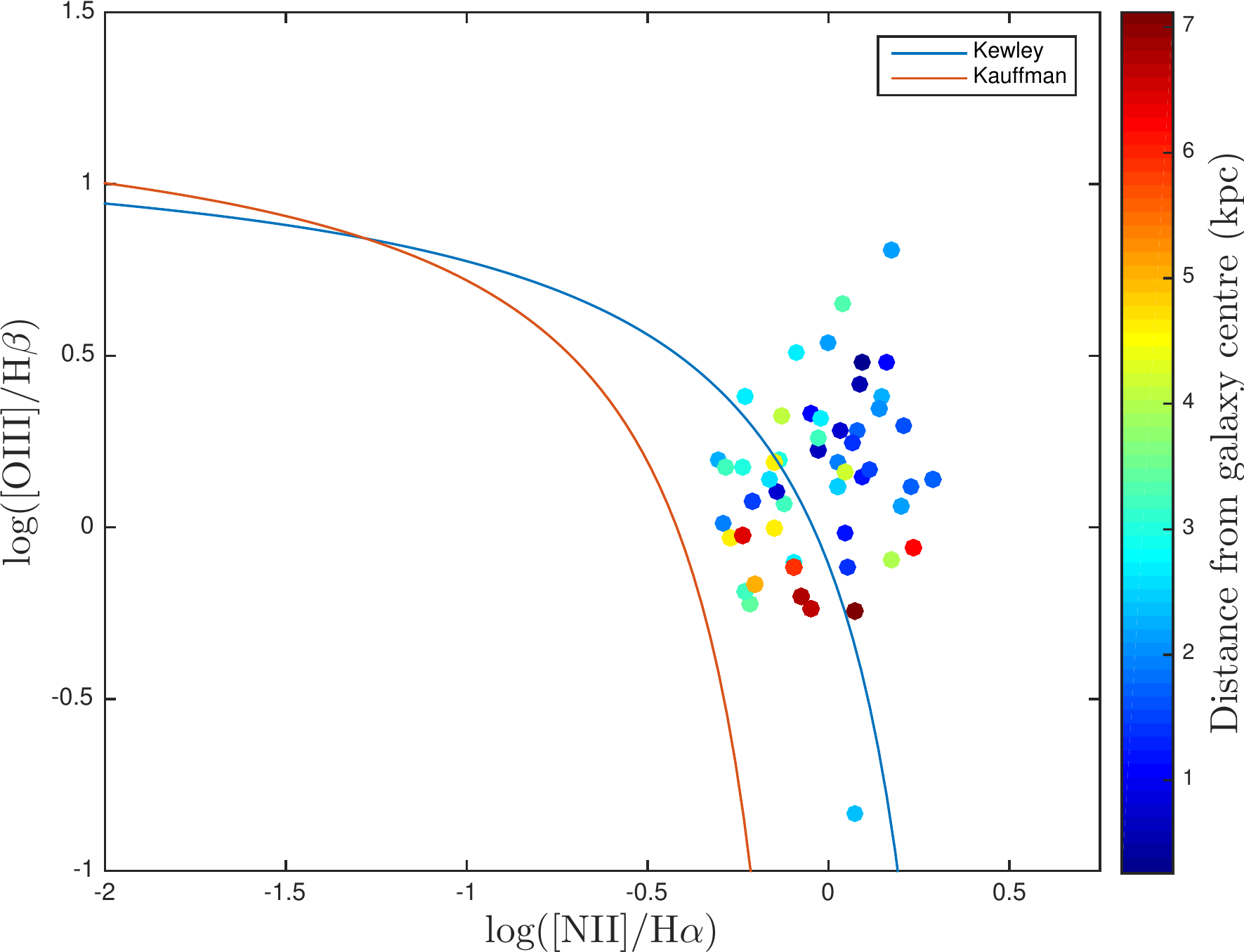}
\caption{IC 1059}
\end{subfigure}
\hfill
\begin{subfigure}{0.3\textwidth}
\centering
\includegraphics[width=\textwidth]{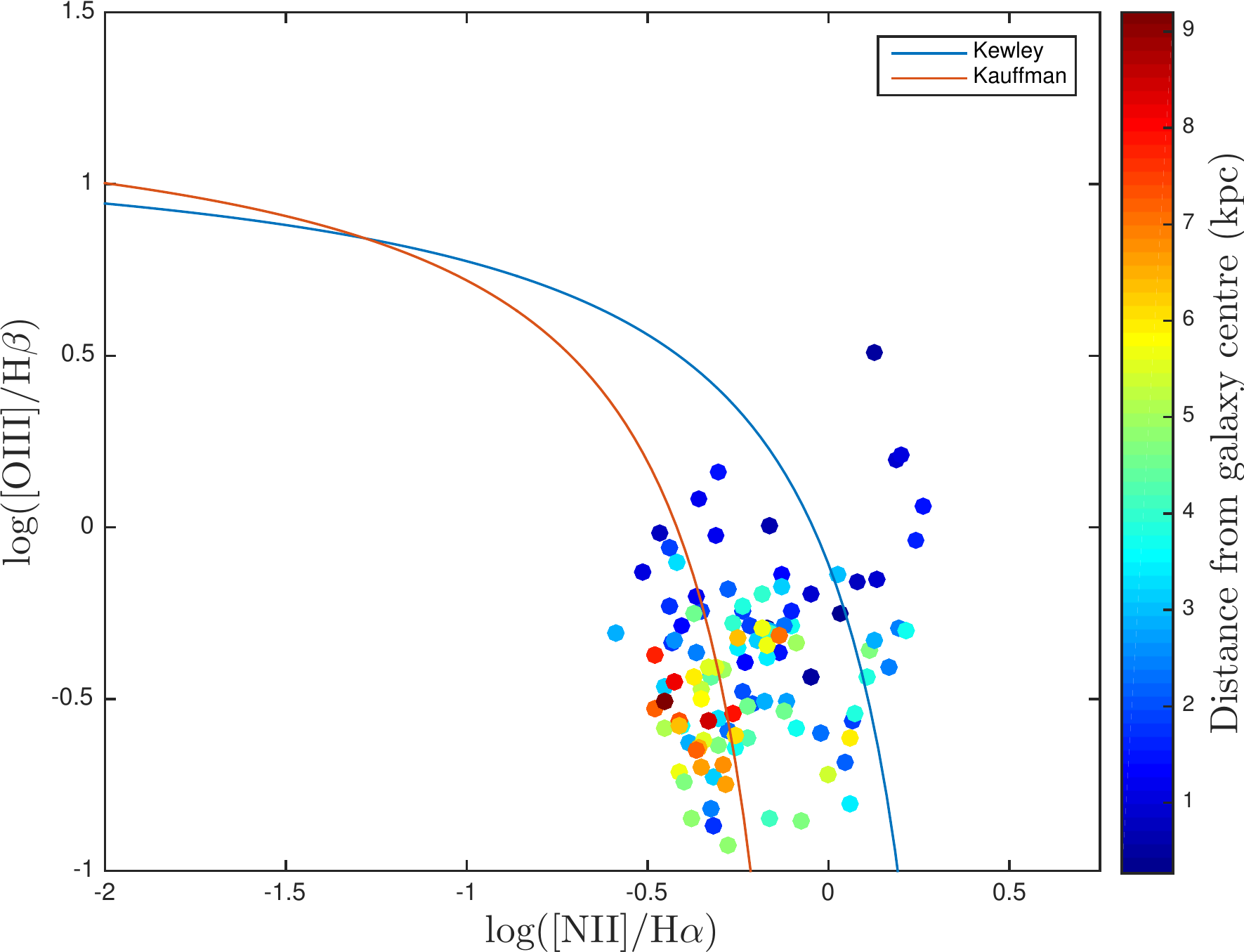}
\caption{IC 0653}
\end{subfigure}
\hfill\begin{subfigure}{0.3\textwidth}
\centering
\includegraphics[width=\textwidth]{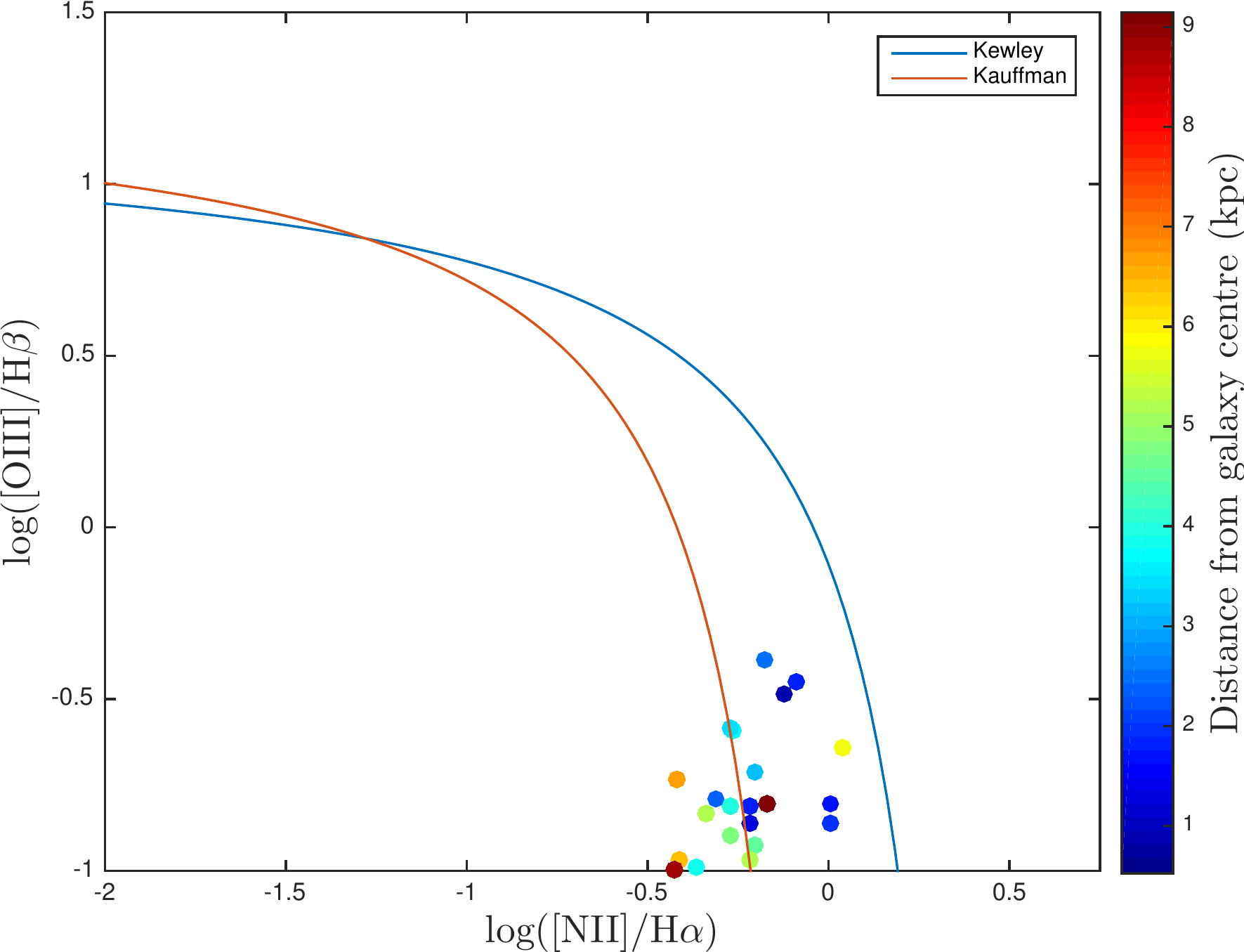}
\caption{CGCG 017-063}
\end{subfigure}
\hfill
\caption{BPT diagrams for sample as a function of distance from galaxy centre for the nine void galaxies studied in this work. The \citet{Kauffmann03} and \citet{Kewley06} lines are in orange and blue respectively. We expect CGCG 019-064 to posses an AGN by the tight correlation of its central points in the AGN region, the rest of the galaxies should be checked against the WHAN diagrams in Figure~\ref{WHAN} for fake AGN.}
\label{BPT}
\end{figure*}

\begin{figure*}
\centering
\begin{subfigure}{0.3\textwidth}
\includegraphics[width=\textwidth]{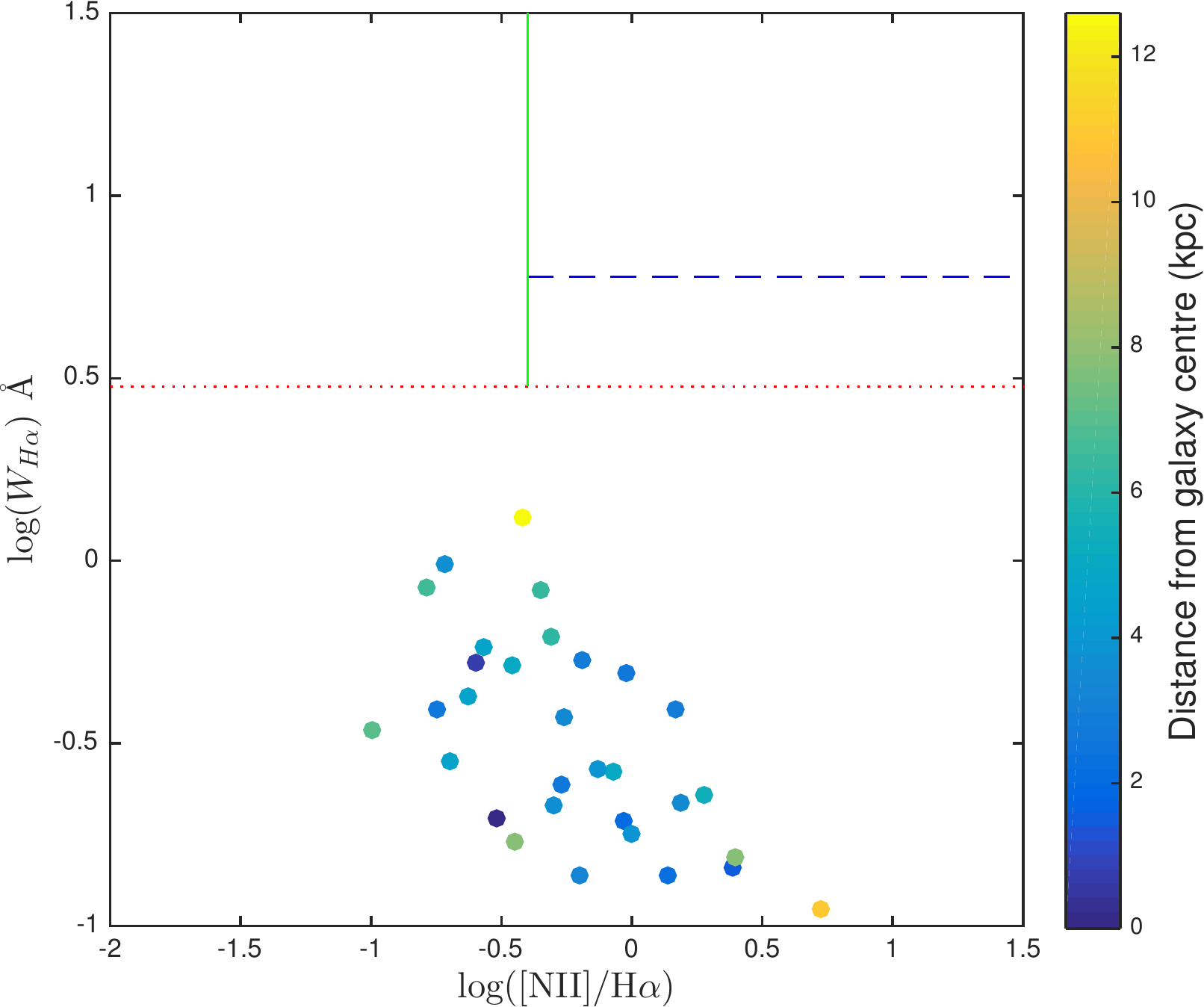}
\caption{IC 0566 }
\end{subfigure}
\hfill
\begin{subfigure}{0.3\textwidth}
\centering
\includegraphics[width=\textwidth]{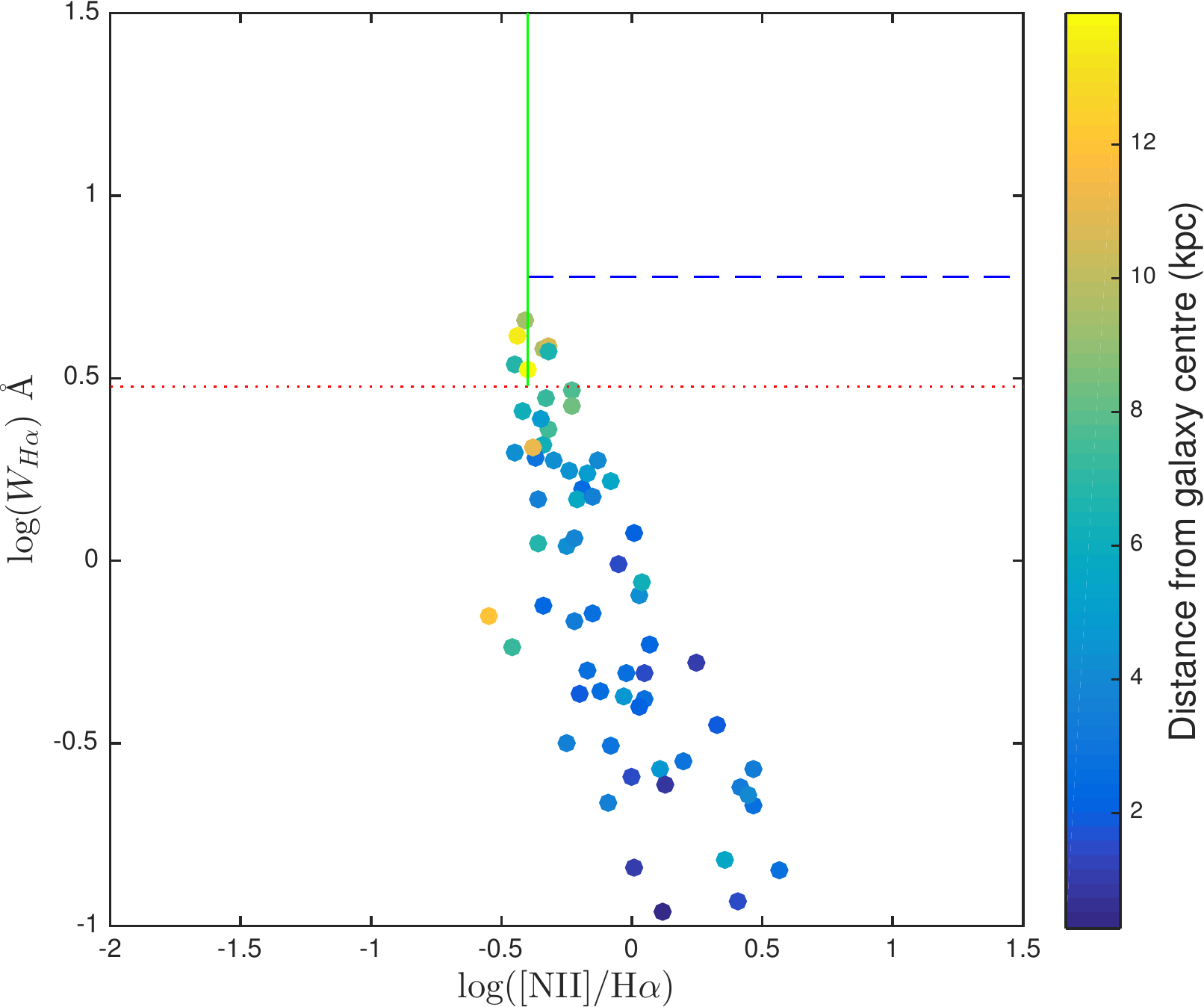}
\caption{CGCG 010-071}
\end{subfigure}
\hfill
\begin{subfigure}{0.3\textwidth}
\centering
\includegraphics[width=\textwidth]{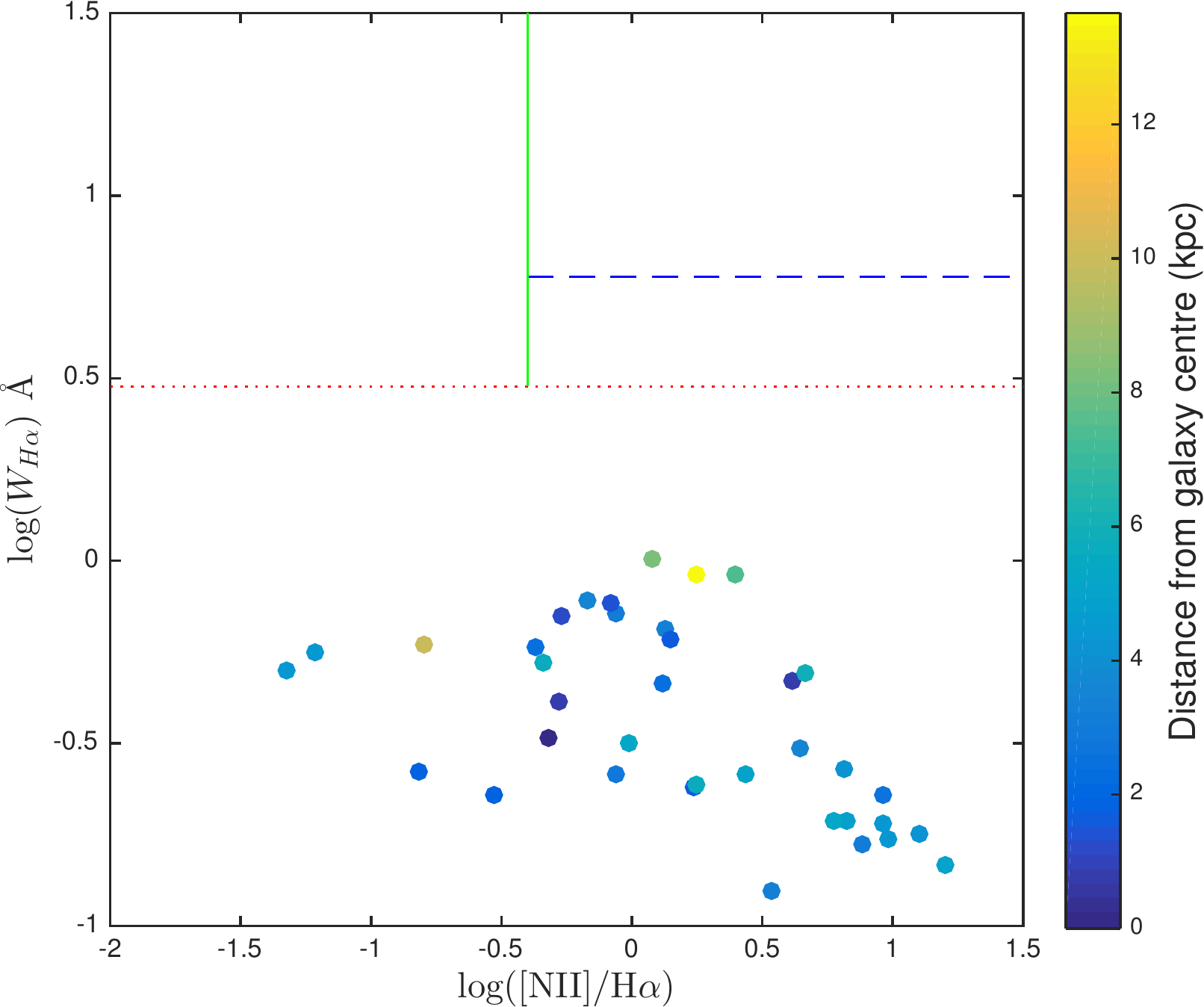}
\caption{GAMA 214363}
\end{subfigure}
\vskip\baselineskip
\begin{subfigure}{0.3\textwidth}
\centering
\includegraphics[width=\textwidth]{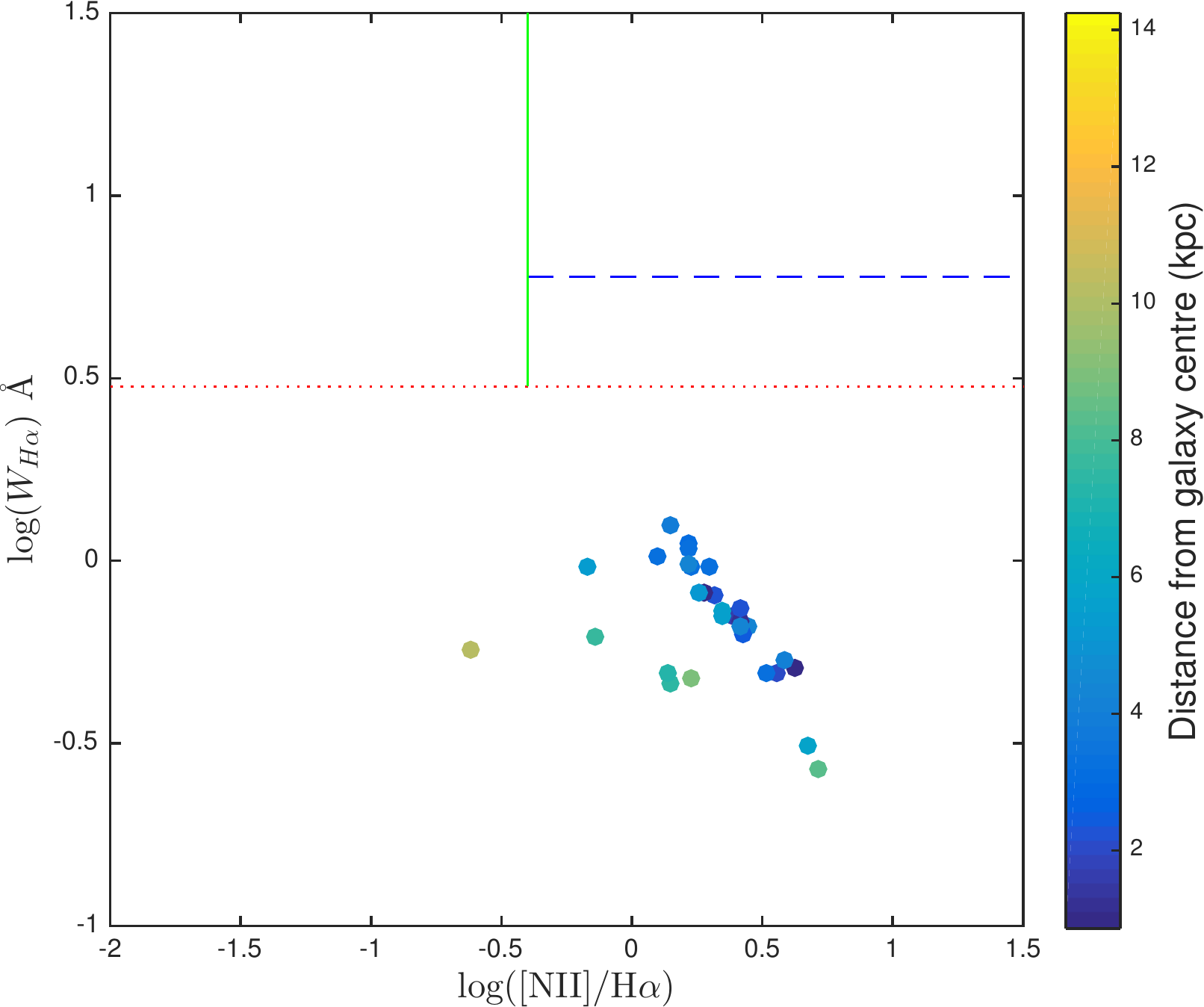}
\caption{GAMA 534760}
\label{no3}
\end{subfigure}
\hfill
\begin{subfigure}{0.3\textwidth}
\centering
\includegraphics[width=\textwidth]{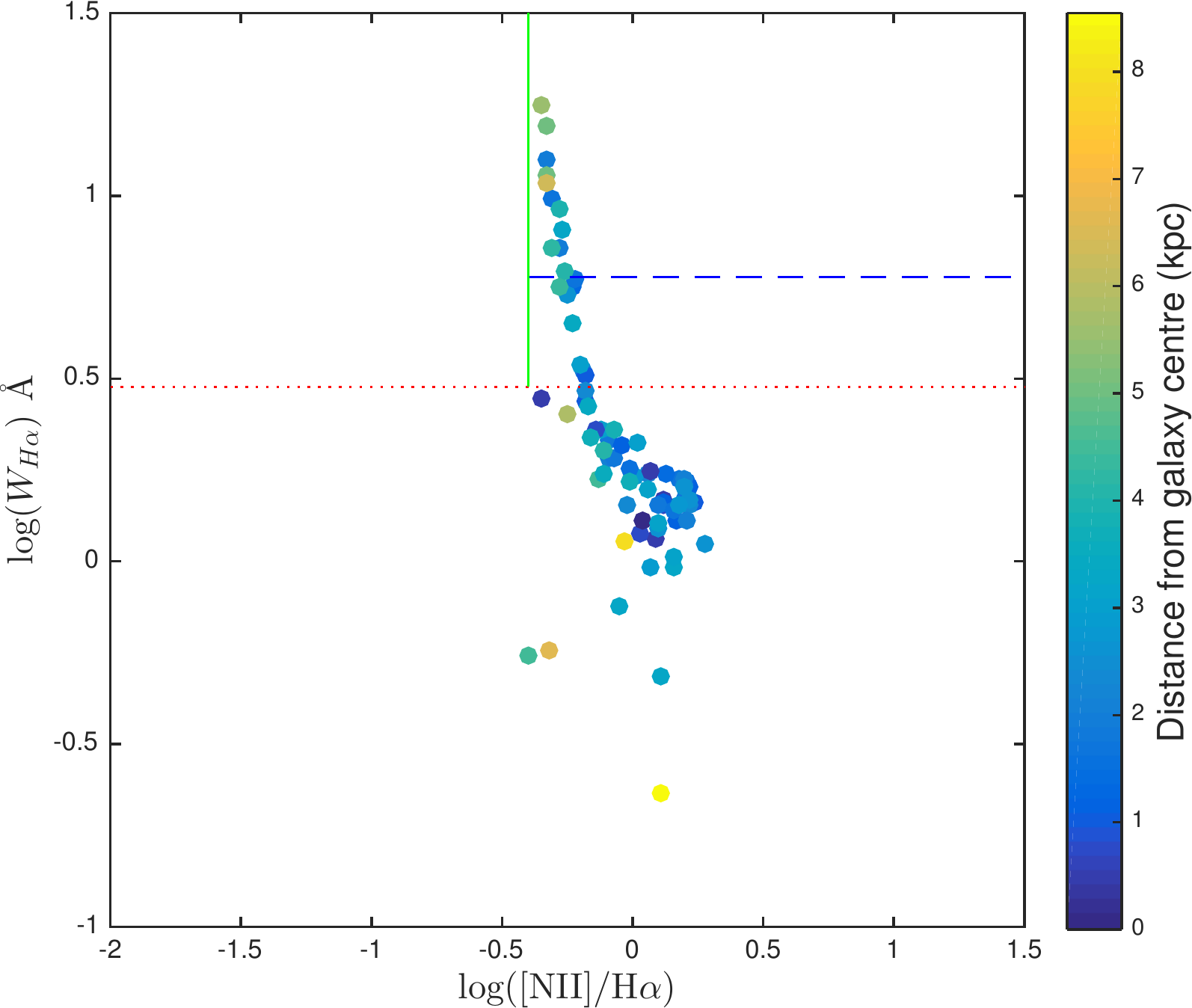}
\caption{CGCG 013-075}
\end{subfigure}
\hfill\begin{subfigure}{0.3\textwidth}
\centering
\includegraphics[width=\textwidth]{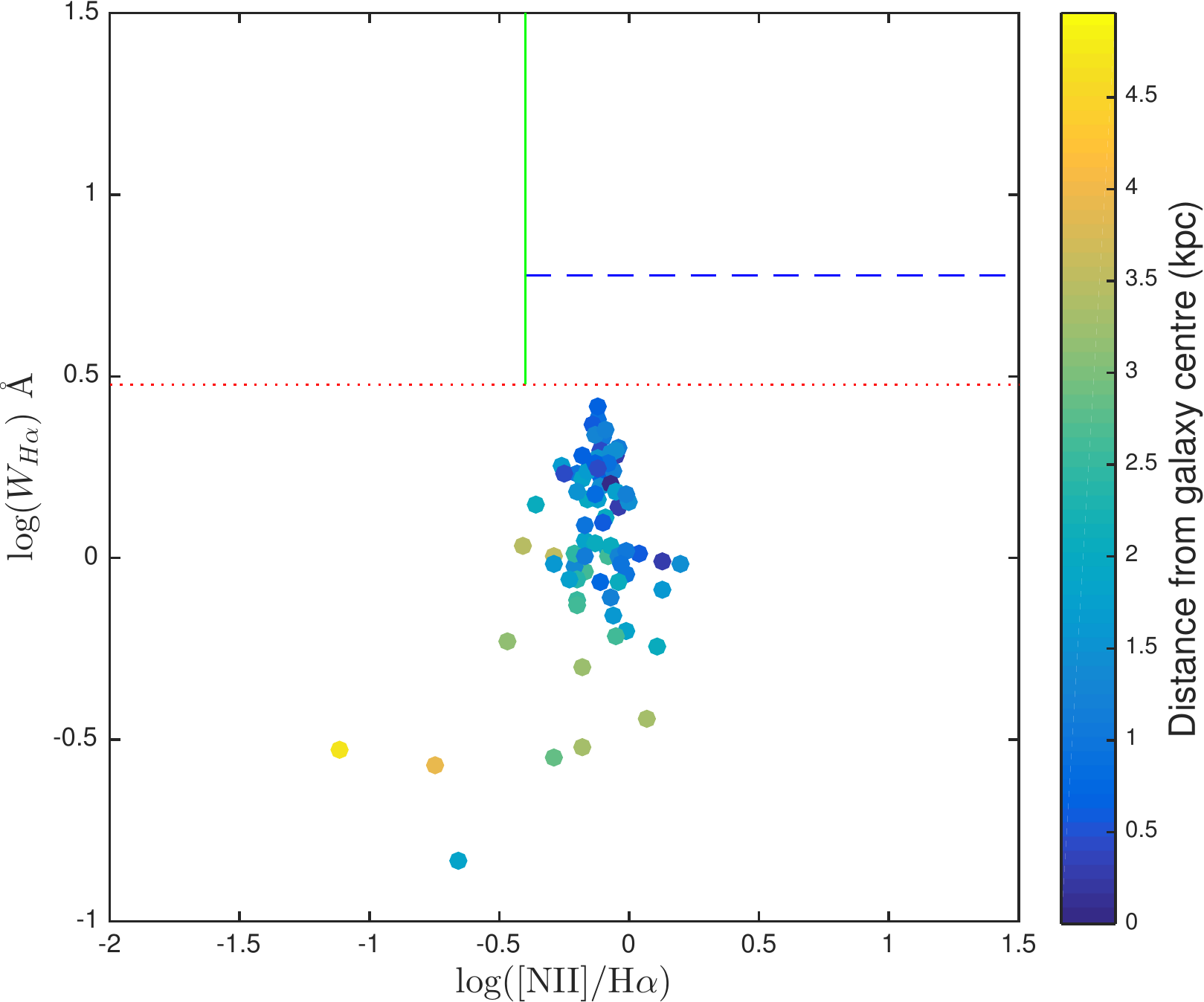}
\caption{CGCG 019-064}
\end{subfigure}
\vskip\baselineskip
\begin{subfigure}{0.3\textwidth}
\centering
\includegraphics[width=\textwidth]{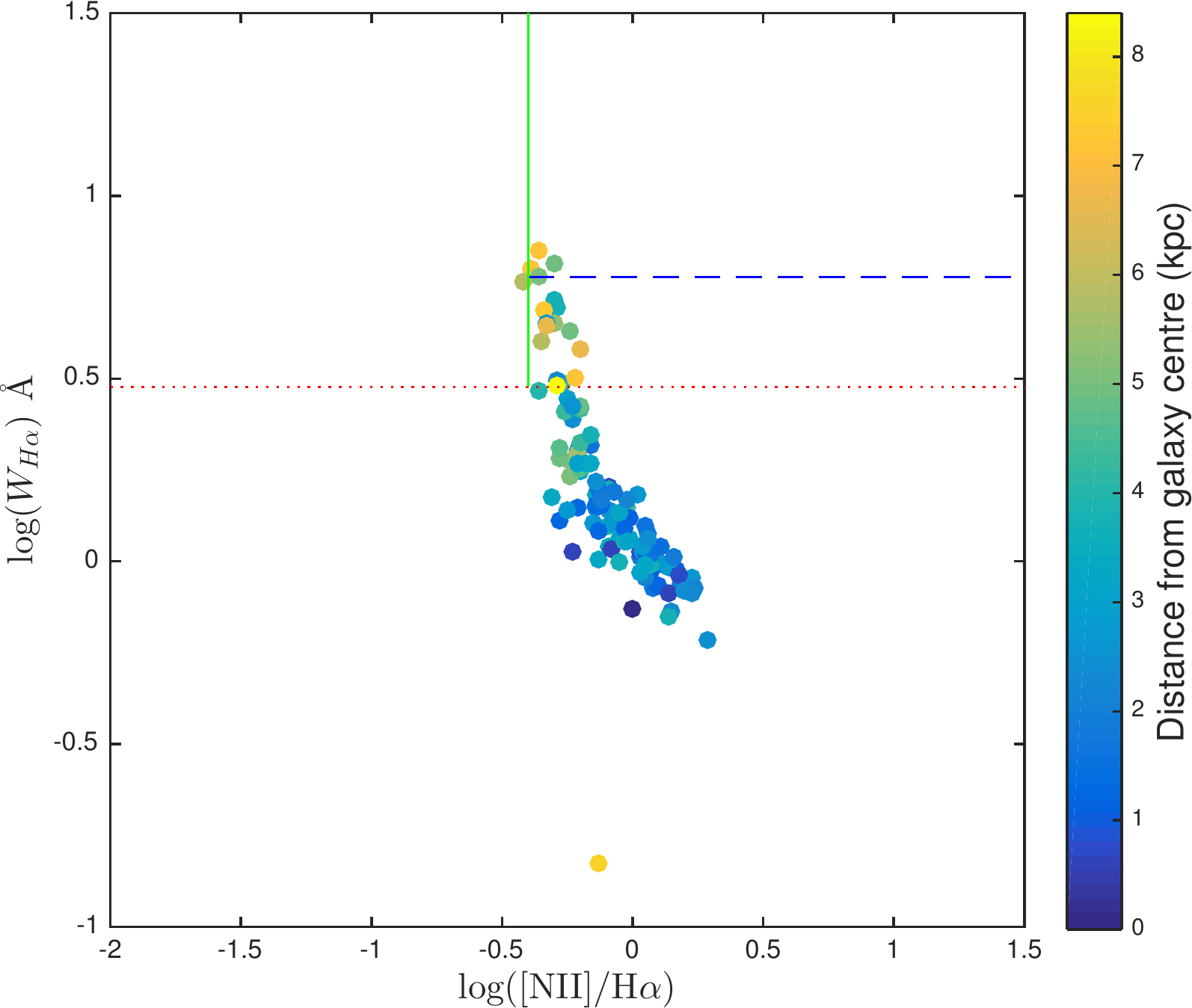}
\caption{IC 1059}
\end{subfigure}
\hfill
\begin{subfigure}{0.3\textwidth}
\centering
\includegraphics[width=\textwidth]{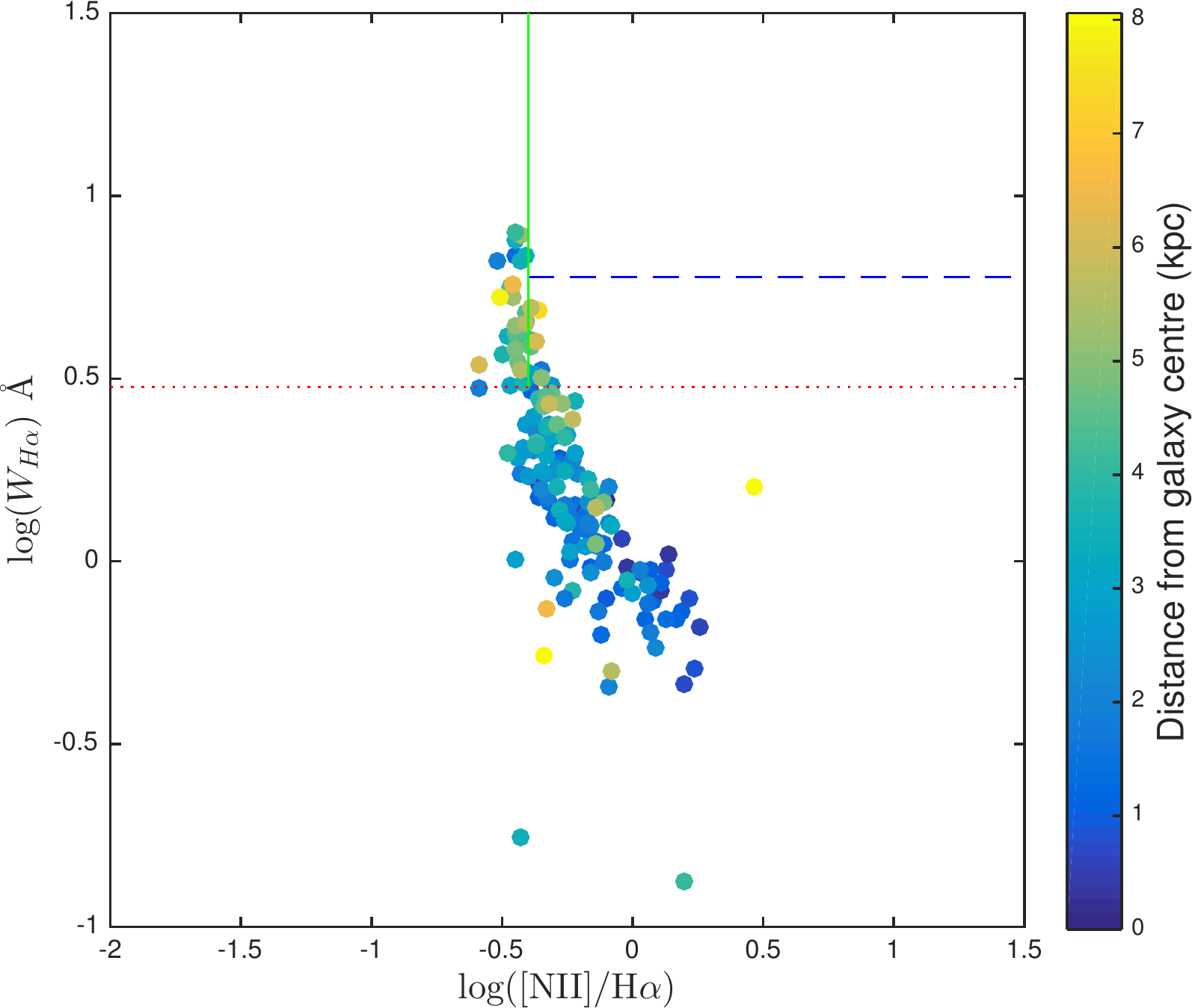}
\caption{IC 0653}
\end{subfigure}
\hfill\begin{subfigure}{0.3\textwidth}
\centering
\includegraphics[width=\textwidth]{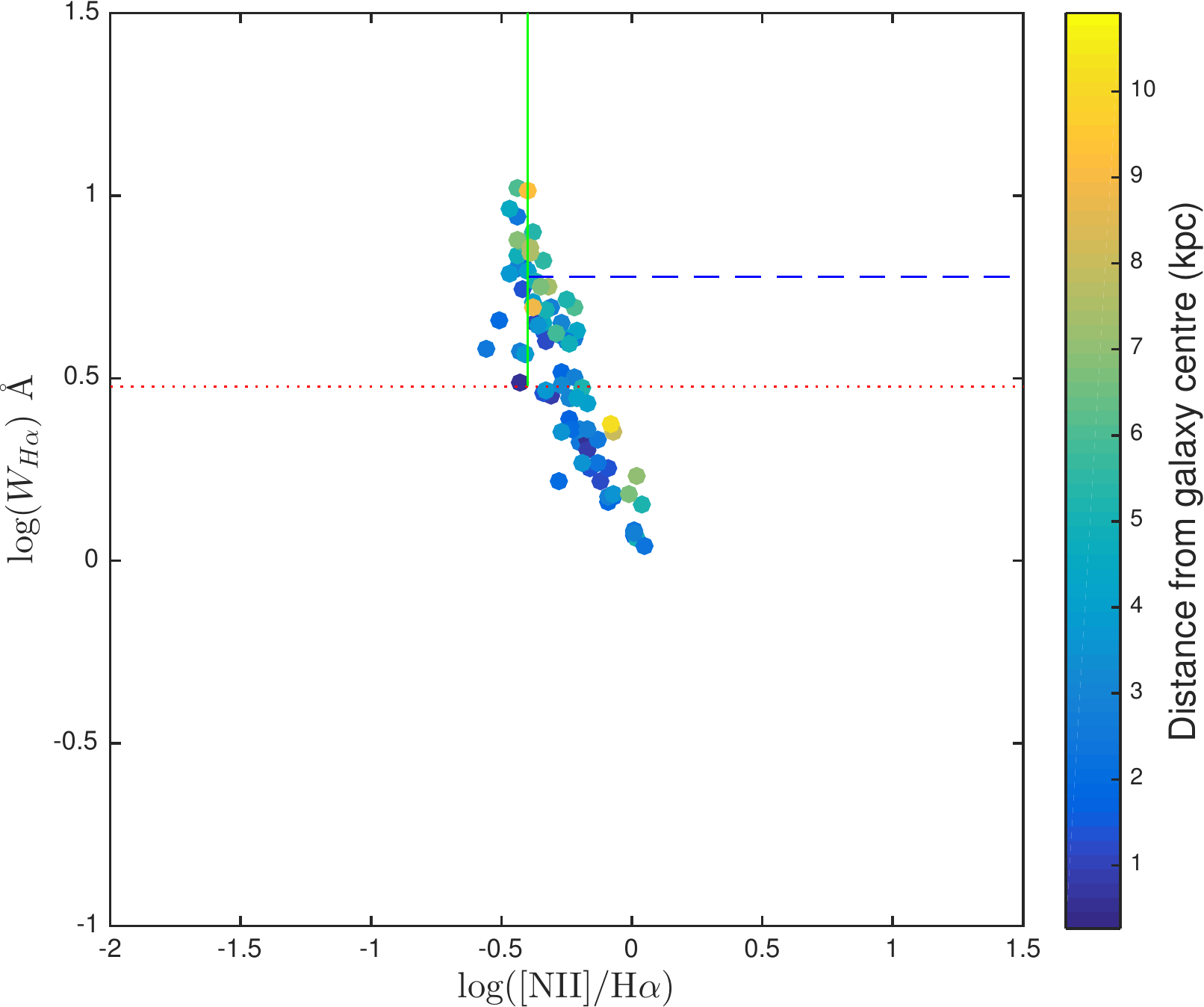}
\caption{CGCG 017-063}
\end{subfigure}
\hfill
\caption{WHAN diagrams for the sample as a function of distance from galaxy centre. Dividing lines are plotted as per \citet{Cid11} to delineate between star forming, passive/ retired and strong and weak AGN regions. Five of the galaxies with high S:N show a tight correlation between $\log(\textrm{[NII]}/\textrm{H}\alpha)$ and log$(\textrm{W}_{H\alpha})$, though only CGCG 013-075 has central points in the AGN region. For this reason, we classify this galaxy as an AGN.}
\label{WHAN}
\end{figure*}

 While we assume that any detected H$\alpha$ is primarily the result of star formation, \citet{Constantin08} showed the prevalence of AGN in void galaxies is similar to that of denser wall regions. We therefore check if any of the H$\alpha$ emission in our galaxies is the result of AGN activity.
In Figure~\ref{BPT} we present Baldwin, Phillips \& Trelivich \citep[BPT;][]{Baldwin81} diagrams and in Figure~\ref{WHAN} we present $W_{H\alpha}$ versus [NII]/H$\alpha$ \citep[WHAN;][]{Cid11} emission line diagnostic diagrams as a function of distance from the galaxy centre. Every point on each diagram corresponds to a Voronoi bin with emission of the relevant lines. Distance to galaxy centre is measured from the flux-weighted centre of the Voronoi bin.  
 The BPT diagrams show most regions of most galaxies residing in either the AGN or AGN+SF regions, though the outer Voronoi bins will suffer from either low S:N, or despite being located in the AGN regions, will be retired \citep[`fake AGN';][]{Cid11}. CGCG019-064 has all its central bins located tightly in the AGN region, and along with the high concentrations of H$\alpha$ and $[\textrm{OIII}]$ in the central regions of this galaxy, we expect it to host an AGN.
 
When used in conjunction with a BPT diagram, a WHAN diagram can separate regions of AGN activity from retired regions of a galaxy with similar $\textrm{[NII]}/\textrm{H}\alpha$. 
Indeed, the central regions of a galaxy located in the AGN region of a BPT diagram should only be classified as a bone fide AGN if they also lie in the AGN region of a WHAN diagram. We see that for the majority of our galaxies, this is not the case.
Of the galaxies with high S:N in Figure~\ref{WHAN}, we see five follow a tight correlation through the star forming region for the outer bins, and into the passive/retired region for the central bins. CGCG 013-075 extends significantly into the AGN region of the WHAN diagram, so we classify it as an AGN. While CGCG 019-064 does not have points in the AGN region of this diagram, given other evidence such as high concentrations of H$\alpha$ and [OIII] in the central regions, we still assume AGN activity, despite the narrow H$\alpha$ equivalent widths.
None of our nine galaxies satisfy the mid-IR strong AGN selection criteria of \citet{Jarrett11}. This, along with a lack of emission line broadening of the spectra lead us to believe these are Seyfert galaxies.
 \citet{Constantin08} show that for the median mass of our sample $(\sim10^{10.5}\textrm{M}_{\odot})$, the fraction of weak AGN expected in void regions is approximately 0.07, while the strong AGN fraction is $\sim$0.05. It is therefore feasible that there are two galaxies hosting AGN in our sample of nine.
 
\section{Comparison to a sample of `Field' galaxies}
\label{ComparisonSection}
We have concluded that despite having predominantly secular evolution histories and similarly low SFRs, our void galaxy sample must still possess a wide range of assembly histories from their kinematics and AGN fraction. We seek to discover how these attributes compare against a sample of galaxies in the wider galaxy population.
We compare the properties of our void galaxy sample to that of a mass, magnitude and redshift-matched sample of field galaxies, drawn from the SAMI galaxy survey early data release \citep[EDR;][]{Croom12,Allen15,Bryant15,Sharp15}.
\begin{table*}
\caption{SAMI EDR galaxies used to create a `field' sample to compare our void galaxies to.}
\label{SAMIcomparison}
\begin{tabular}{l c c c c l c c }
\hline
\textbf{Galaxy} & \textbf{Redshift} $^{1}$& \textbf{$m_{r}^{2}$} &$u-r^{3}$ & \textbf{Stellar mass}$^{4}$  & \textbf{Environment}$^{5}$ & \textbf{SFR}$^{6}$ & \textbf{WISE SFR}$^{7}$ \\
                        &                                      &   &                 &($\textbf{M}_{}\odot$)   &   & ($\textbf{M}_{}\odot~\textbf{yr}^{-1}$) & ($\textbf{M}_{}\odot~\textbf{yr}^{-1}$) \\
\hline
GAMA 47342  & 0.024 & 15.36 & 2.01 & 10.07 & Isolated & 1.62 & -\\
GAMA 79635  & 0.040 & 15.05 & 2.16 &10.45 & Isolated  & 1.79 & 1.70\\
GAMA 91924  & 0.052 & 15.04 & 2.25 &10.61 & Galaxy pair member  & 2.55 & 1.96\\
GAMA 91963  & 0.050 & 14.52 & 2.61 &11.02 & Galaxy pair member & 0.93 & 0.77\\
GAMA 230714 & 0.024 & 14.95 & 2.25 &10.19 & Group member (n=35) & 1.81 & 1.99\\
GAMA 517302 & 0.029 & 15.29 & 2.18 &10.21 & Group member (n=21) & 1.62 & 0.46 \\
GAMA 536625 & 0.037 & 15.54 & 2.21 &10.21 & Isolated & 0.90  & -\\
GAMA 570206 & 0.043 & 15.60 & 2.78 &10.51 & Group member (n=19) & 0.96 & 0.29 \\  
GAMA 599582 & 0.052 & 14.84 & 2.71 &10.75 & BCG of group (n=3) & 1.76  & 2.18\\

\hline
 \multicolumn{7}{l}{$^{1}$GAMA} \\
 \multicolumn{7}{l}{$^{2}$GAMA r-band model magnitudes} \\
 \multicolumn{7}{l}{$^{3}$ SDSS model magnitudes.}\\
 \multicolumn{7}{l}{$^{4}$ From \citet{Taylor11}}\\
 \multicolumn{7}{l}{$^{5}$  From \citet{Robotham11}. }\\
 \multicolumn{7}{l}{$^{6}$ Using the relation of \citet{Richards15}} \\
 \multicolumn{7}{l}{$^{7}$ 12$\mu$m SFR relation from \citet{Cluver14} adjusted from a \citet{Salpeter55} IMF using the conversion  }\\
 \multicolumn{7}{l}{of \citet{Gunawardhana13}.}
\end{tabular}
 \end{table*}
 
 \subsection{Field Galaxy Sample}
\begin{figure*}
\centering
\begin{subfigure}{0.3\textwidth}
\includegraphics[width=\textwidth]{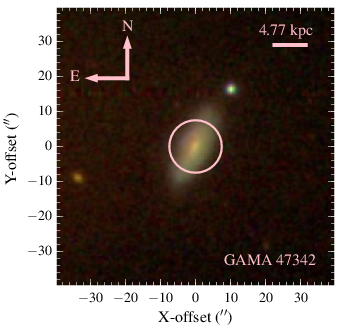}
\caption{GAMA 47342 }
\end{subfigure}
\hfill
\begin{subfigure}{0.3\textwidth}
\centering
\includegraphics[width=\textwidth]{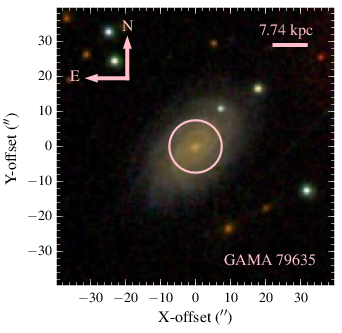}
\caption{GAMA 79635}
\end{subfigure}
\hfill
\begin{subfigure}{0.3\textwidth}
\centering
\includegraphics[width=\textwidth]{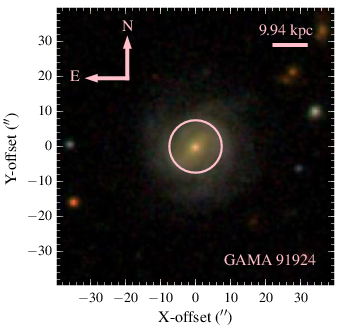}
\caption{GAMA 91924}
\end{subfigure}
\vskip\baselineskip
\begin{subfigure}{0.3\textwidth}
\centering
\includegraphics[width=\textwidth]{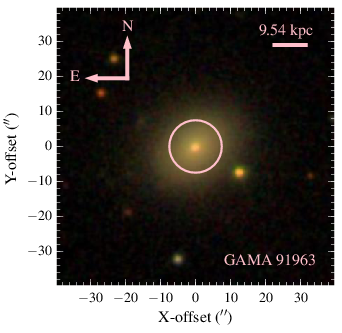}
\caption{GAMA 91963}
\end{subfigure}
\hfill
\begin{subfigure}{0.3\textwidth}
\centering
\includegraphics[width=\textwidth]{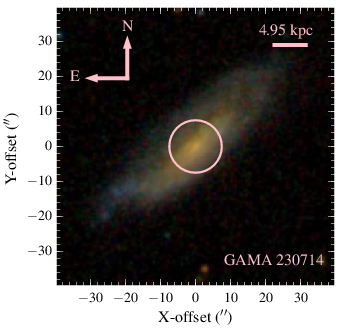}
\caption{GAMA 230714}
\end{subfigure}
\hfill\begin{subfigure}{0.3\textwidth}
\centering
\includegraphics[width=\textwidth]{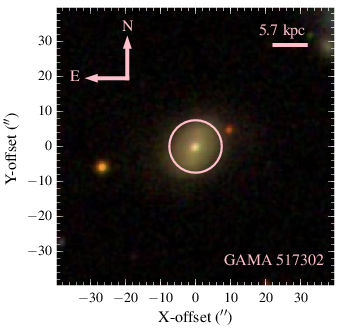}
\caption{GAMA 517302}
\end{subfigure}
\vskip\baselineskip
\begin{subfigure}{0.3\textwidth}
\centering
\includegraphics[width=\textwidth]{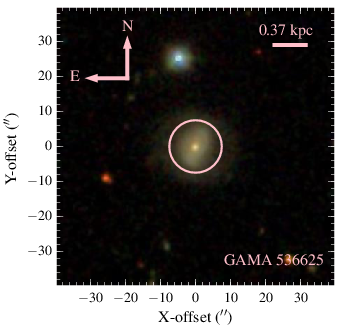}
\caption{GAMA 536625}
\end{subfigure}
\hfill
\begin{subfigure}{0.3\textwidth}
\centering
\includegraphics[width=\textwidth]{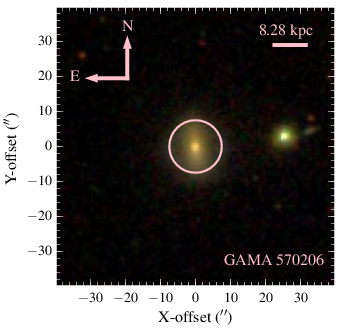}
\caption{GAMA 570206}
\end{subfigure}
\hfill\begin{subfigure}{0.3\textwidth}
\centering
\includegraphics[width=\textwidth]{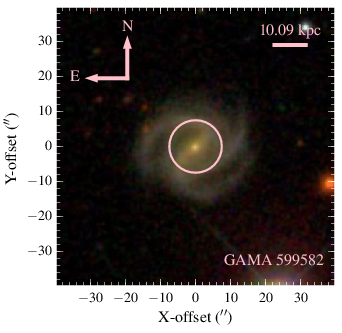}
\caption{GAMA 599582}
\end{subfigure}
\hfill
\caption{SDSS colour images of the nine SAMI EDR field galaxies in the sample with the SAMI field of view overlaid in pink.}
\label{SAMI_gals}
\end{figure*}

 A comparison sample of field galaxies taken from all environments was created from the SAMI galaxy survey EDR by matching to the constraints used to select our void  galaxy targets. All galaxies with stellar mass $>10^{10}\textrm{M}_{\odot}$, redshift $<0.052$ and r-band apparent magnitude $<15.7$ were included in the sample. This selection ensures the galaxies we are comparing to are similarly massive.  
 Of the 107 galaxies in the SAMI EDR, nine fit these criteria, the properties of which are described in Table~\ref{SAMIcomparison}, with SDSS colour images in Figure~\ref{SAMI_gals}. Despite the identical selection method, these nine galaxies are bluer than the void sample, spread throughout the star forming main sequence and quiescent regions. 
 On a $u-r$ colour magnitude diagram using SDSS photometry, four galaxies lie on the red sequence, four in the green valley, and one in the blue cloud. 
 Their mid-IR colours show a slightly higher excess of $W3$ than the void sample of this work. 
 We check that these field galaxies are in sufficiently denser regions of the cosmic web than our void galaxies by determining the distance to the fifth nearest neighbour ($d_{5}$) and  the environmental density ($\sigma_{5}$) of the field sample to compare to the void sample.  We use the SDSS DR7 sample with an absolute magnitude cut of $\textrm{M}_{R}>-20$ to determine $d_{5}$ and $\sigma_{5}$ following the method described in \citet{Baldry06}.
 Our field galaxies have a median $d_{5}=1.8~\textrm{Mpc}$ and $\sigma_{5}=0.504~\textrm{Mpc}^{-2} $, compared to 4.8 Mpc  and 0.069 $\textrm{Mpc}^{-2}$ for void regions respectively. \citet{Baldry06} states that void regions should have $\sigma_{5}$ as low as 0.05 $\textrm{Mpc}^{-2}$, increasing to 20 $\textrm{Mpc}^{-2}$ for cluster regions for SDSS DR4. We are therefore satisfied our sample galaxies are located in bona fide void regions, and are sufficiently more underdense then the comparison field galaxies.

  At masses $>10^{10}~\textrm{M}_{\odot}$, mass quenching plays an increasingly important role in galaxy evolution \citep[e.g.,][]{Geha12}, hence we choose to match our void and field sample based on mass rather than environment. Because of this, there may be some secondary environmental effects, for example those caused by nearby neighbours, but with our small sample, we cannot investigate this in detail. Instead, we seek to gain a representative idea of the global properties of isolated void galaxies as compared to the rest of the Universe. 
 We compare these galaxies to our original void galaxy sample in the following sections.

\subsection{Morphology \& Star Formation Rates}
The SDSS colour images of the nine SAMI field galaxies are shown in Figure~\ref{SAMI_gals}. These galaxies all appear discy, with a mix of spiral and lenticular morphologies. This is a similar spread to the void galaxy sample. \\

 \begin{figure*}
 \centering
 \begin{subfigure}{0.45\textwidth}
 \includegraphics[width=\textwidth]{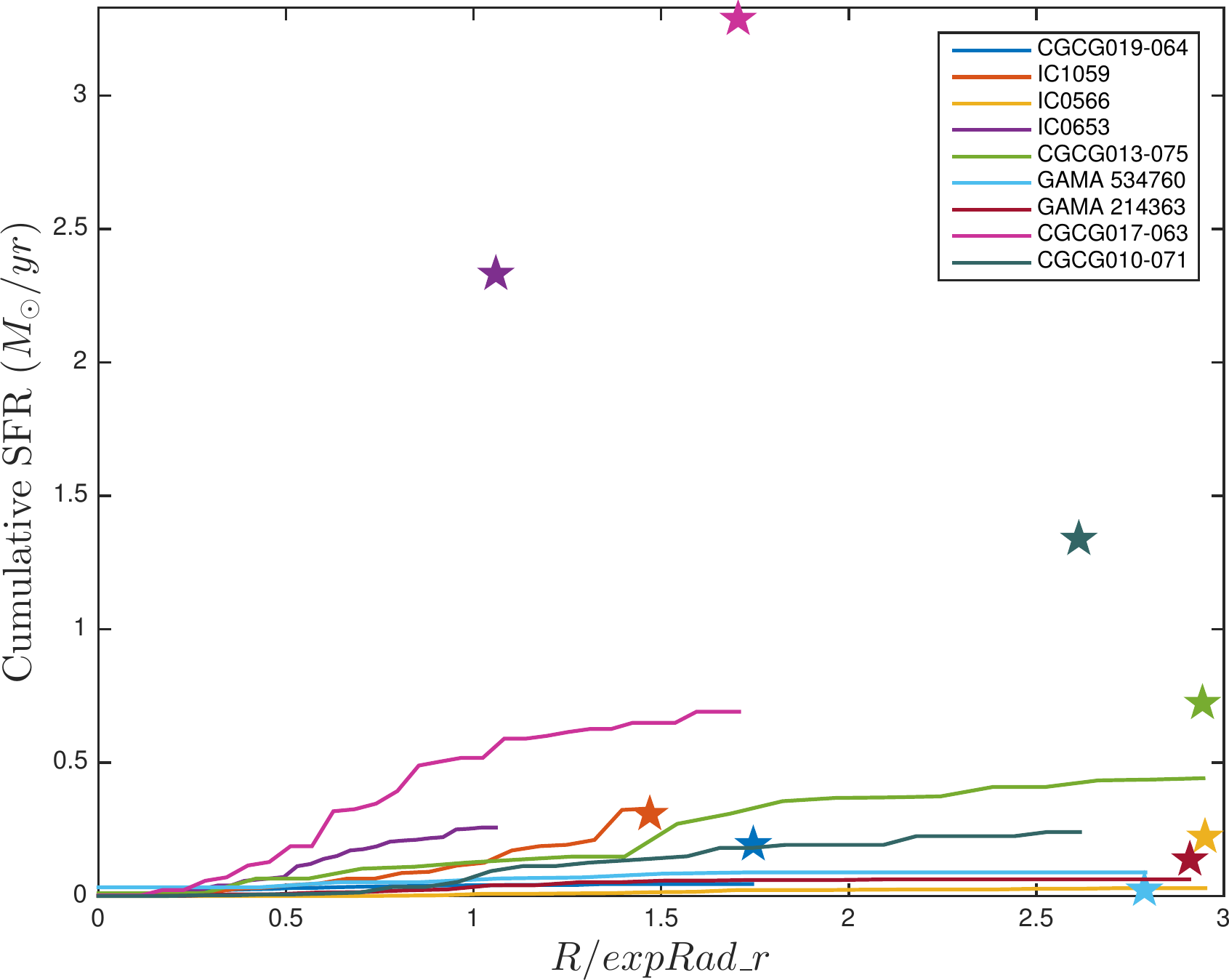}
\caption{Void galaxy sample. }
\end{subfigure}
\hfill
 \begin{subfigure}{0.45\textwidth}
 \includegraphics[width=\textwidth]{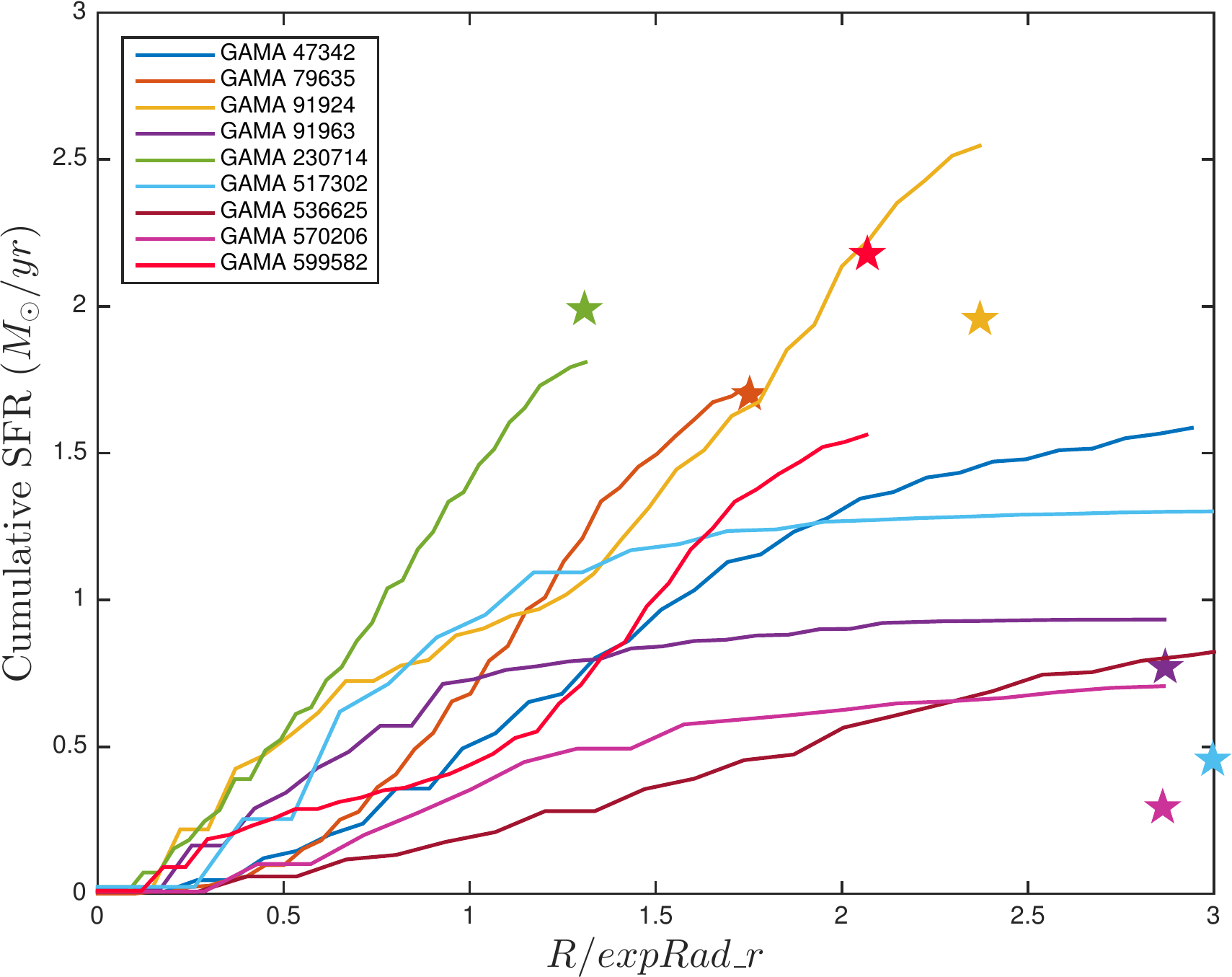}
\caption{Field galaxy comparison sample. }
\end{subfigure}
\hfill
\caption{Cumulative H$\alpha$-derived star formation rates for the nine galaxies in the void galaxy sample and the SAMI comparison sample (coloured lines) and WISE $W3$ total integrated SFRs (stars) where available . All integrated star formation rates are calculated to be $< 1 \textrm{M}_{\odot}~\textrm{yr}^{-1}$ for the void galaxy sample, and $< 3 \textrm{M}_{\odot}~\textrm{yr}^{-1}$ for the SAMI sample, truncated at $3\times expRad_r$ where required. While on average, the field galaxy integrated SFRs are higher, flux calibration issues and disc regions missed due to low S:N of the void galaxy sample should be taken into account. The two galaxies with the highest discrepancies between H$\alpha$ and WISE $W3$ SFRs were only observed out to 1 and 1.5 $expRad_r$. The superior S:N of the SAMI galaxies means they are often measured out past 2 $expRad_r$.
When instead the WISE 12$\mu$m SFRs are compared, both samples have a very similar spread in SFR, leading us to conclude there is no significant difference between the void and field sample SFRs..}
\label{SAMI_SFR}
\end{figure*}

We present the H$\alpha$-derived cumulative star formation rates as a function of radius in Figure~\ref{SAMI_SFR}, calculated in the same manor as for the void galaxies using Equation~\ref{eq1}, but without the need for Voronoi binning of flux due to the superior S:N of the SAMI data. Also included as star-shaped points of the same colour are WISE integrated SFRs for each galaxy when available.
While we would expect from prior literature that void galaxies would be on average more star forming than field galaxies, we find this is not the case. The low S:N of the outer regions of the void galaxies may be contributing to this low SFR however.
On average, the SAMI field galaxies posses SFRs greater than that of the void sample. This is not unexpected, as the field galaxies had on average a greater excess of mid-IR W3 emission. 
If we instead compare WISE SFRs, these galaxies have more comparable SFRs. The highest SFR belongs to a void galaxy, 
though on average field galaxies still have slightly higher SFRs than void galaxies.
We also note that while there are examples of field galaxies with high SFRs and close companions (as seen in Table~\ref{SAMIcomparison}), we cannot conclude that star formation is enhanced by the companions, as we also see examples of high star formation in isolated galaxies. While it is difficult to separate the contribution of large-scale environmental effects and small-scale neighbour effects on star formation, the fact that both high and low star forming galaxies are seen in all environments in the field sample leads us to believe that neither are having a significant effect on our small field sample.

Though our samples of both void and field galaxies are not large enough to comment statistically on the average SFRs of each environment, we see no evidence for an enhancement of star formation in void galaxies for this mass range.
Since these galaxies are mass-matched, we infer that their relative environments have little effect on star formation, instead quenching is occurring based on the galaxy's mass.

  \section{Conclusions}
  To investigate the assembly histories of void galaxies, we took IFS observations of a sample of nine massive ($\textrm{M}_{\star}>10^{10}\textrm{M}_{\odot}$) central and isolated void galaxies.
  While gas and stellar kinematics revealed seven of our nine galaxies are regularly rotating as disc galaxies, the remaining two posses gas kinematics consistent with a galaxy that has either undergone a recent merger or gas mixing due to the presence of a galactic bar. Visual morphologies confirm these two diagnoses, and verify the rest of the sample are indeed secularly-evolving disc galaxies. 
  Two of our galaxies are located on BPT and WHAN diagnostic diagrams consistent with AGN activity. This range of galaxy properties for a sample of galaxies of similar mass and environment confirms the results of \citet{Penny15} that there must be many assembly histories present in void galaxies, and highlights the interactions still occurring in voids \citep[e.g.,][]{Kreckel12}.
  Despite the mix of internal processes occurring in these galaxies, all have uniformly low star formation rates of $<1\textrm{M}_{\odot}~\textrm{yr}^{-1}$ from the central regions of integrated H$\alpha$-derived SFRs, or $<3.3~\textrm{M}_{\odot}~\textrm{yr}^{-1}$ using the WISE mid-IR photometric relation of \citet{Cluver14}. 
  
  We compared our void galaxy sample to that of a mass, magnitude and redshift-matched sample of field galaxies from the SAMI EDR. We find similarly low SFRs for both samples, and no evidence of an enhancement of SFR due to the relative isolation of the void sample. Indeed, some of our void galaxies have lower SFRs than the field sample. We are led to conclude that the environmental position of these galaxies is not affecting their SFRs, and that mass-quenching is the dominant mode of star formation cessation in galaxies of this mass range.
\appendix
\section{Notes on Individual Void Galaxies Observed}
We discuss each galaxy observed for this work in detail.
\subsection{IC 0566}
This galaxy has an obvious merger remnant in the tidal tail located in the South West corner. It is classified as lenticular by \citet{Nair10}. The negligible H$\alpha$ emission at the centre of the galaxy coupled with a low star formation rate of $0.03~\textrm{M}_{\odot}~\textrm{yr}^{-1}$ implies the recent merger was dry. Gas kinematics show that the galaxy has not yet virialised after this merger, though the regular rotation of the stellar component suggests this was a minor merger. While it seems this galaxy is not evolving in isolation, the stellar rotation curve implies until recently, this galaxy was discy. 

\subsection{CGCG 010-071}
\citet{Nair10} classify this galaxy as an edge-on, barred Sa galaxy with a T-type of 1. The stellar velocity map shows regular rotation, and while the gas is still rotating in the same general direction, it is disturbed slightly in the central region, likely by the bar structure. Balmer and forbidden emission line ratio maps show a higher fraction of line emission in the disc region, with lower levels in the central bulge region. The low star formation rate of $0.24~\textrm{M}_{\odot}~\textrm{yr}^{-1}$ lead us to classify this galaxy as a secularly-evolving red spiral.

\subsection{GAMA 214363}
This galaxy has no published classifications, but  using Sloan imaging we classify this galaxy as lenticular. H$\alpha$ and other emission lines have little to no emission in the central region of the galaxy. Gas and stellar kinematics line up well for this galaxy, and along with its low SFR and disc nature, we again classify it as a secularly-evolving disc galaxy.

\subsection{GAMA 534760}
At z=0.052, this galaxy is the highest redshift galaxy in our sample. \citet{Fukugita07} classify this galaxy as lenticular. 
Line emission is concentrated in the outer regions of the galaxy, and regular rotation lines up both the gas and stellar kinematics. Given the slightly higher redshift, but similar stellar mass of this galaxy, it seems to be comparably more evolved compared to its discy counterparts. 

\subsection{CGCG 013-075}
This galaxy is a face-on, disc-dominated, barred spiral galaxy. This galaxy possesses blue spiral arms, and the SFR is low in the inner regions, then increases out past an exponential radius, highlighting the importance of studying a galaxy out past its effective radius. Stellar and gas kinematics match up well, indicating no recent merger activity. BPT and WHAN diagrams both agree that this galaxy is likely to be hosting an AGN. This, coupled with the moderate SFR means this is the most active galaxy in our void sample. It is possible the star formation in the spiral arms is being fuelled by infalling gas perhaps from a filament or tendril extending into the void. 

\subsection{CGCG019-064}
This is the only galaxy with a higher relative H$\alpha$ concentration in the core region than the outskirts.
The emission line flux maps follow that of the H$\alpha$ map, showing a higher concentration of every emission line besides H$\gamma$ in the central regions of the galaxy, unlike the rest of our sample. The high $[\textrm{OIII}]$ concentration in the centre is indicative of AGN activity, and while the BPT diagram reinforces this, the WHAN diagram places the galaxy in the retired region by virtue of the small H$\alpha$ EWs. Despite this, we expect this galaxy to posses a (likely weak) AGN.

\subsection{IC 1059}
This galaxy is another archetypal example of a discy void galaxy. This galaxy comprises a red bulge region, with an outer disc of younger stellar population.
NASA Sloan Atlas\footnote{\url{http://www.nsatlas.org/}} r-band Sersic fit residuals show a ring surrounding the nucleus, possibly the result of a minor merger. The Sersic number is 3.9 and we classify this galaxy as a lenticular.
The stellar and gas kinematics match up well as shown in Figure~\ref{rot_curve}, indicating no recent mergers. The characteristic emission line ratios for Balmer lines, [NII] and [OIII] show low levels of emission in the centre, and higher rings of emission surrounding, all reminiscent of a galaxy with a red and passive centre, but a younger, more active outer region. The symmetry of the emission line ring surrounding the nucleus suggests infalling gas as the mechanism for star formation.

\subsection{IC 0653}
This galaxy has the largest angular size in our sample, so we probe the inner regions in extreme detail, out to half an exponential fit radius. The morphology of this galaxy is again a red spiral, with an inclined axis ratio. The emission line maps have the best S:N of our sample and again show less emission in the central regions compared to the outskirts. We measure a SFR of $0.26~\textrm{M}_{\odot}~\textrm{yr}^{-1}$, which would likely be higher and closer to the WISE mid-IR value if we had probed out further into the disc regions of the galaxy. A WHAN diagram confirms the central regions of the galaxy are passive, with more star forming regions being located towards the outskirts of the disc. Gas and stellar kinematics match up very well, implying no recent mergers. This, coupled with the visible spiral arms and lack of star formation, lead us to classify this galaxy as a red spiral, secularly evolving in isolation.

\subsection{CGCG 017-063}
This galaxy is a face-on spiral galaxy with properties typical of the rest of the void galaxy population. Emission line maps show lower levels of H$\alpha$ emission in the central regions of the galaxy compared to the outer disc. This is true also for the other Balmer lines, [OIII] and [NII]. A BPT diagram places the central spaxels in the AGN+SF region and the outer regions in the purely star forming region We deduce the likelihood of AGN activity is low in this galaxy. This is confirmed by the WHAN diagram, which places the central regions of the galaxy in the passive/retired region, with the outskirts again into the star forming region. We observe this galaxy out to 5 exponential fit radii, with a total integrated H$\alpha$ SFR of $0.70~\textrm{M}_{\odot}~\textrm{yr}^{-1}$. We conclude this galaxy is an anaemic red spiral. 

  \section*{Acknowledgements} 
  The authors wish to thank the referee, whose thoughtful comments and suggestions have improved the quality of this work.
  They also wish to thank John Stott and Mehmet Alpaslan for helpful discussions on this work. The authors are grateful for the help of Adam Schaefer and Scott Croom with the SAMI EDR data, and Nicola Pastorello for initial software guidance.
  AFM acknowledges support from an Australian Postgraduate Award (APA), and a J. L. William postgraduate award.
  SJP acknowledges postdoctoral funding from the University of Portsmouth.\\
 GAMA is a joint European-Australasian project based around a spectroscopic campaign using the Anglo-Australian Telescope. The GAMA input catalogue is based on data taken from the Sloan Digital Sky Survey and the UKIRT Infrared Deep Sky Survey. Complementary imaging of the GAMA regions is being obtained by a number of independent survey programmes including GALEX MIS, VST KiDS, VISTA VIKING, WISE, Herschel-ATLAS, GMRT and ASKAP providing UV to radio coverage. GAMA is funded by the STFC (UK), the ARC (Australia), the AAO, and the participating institutions. The GAMA website is http://www.gama-survey.org/.\\
 The SAMI Galaxy Survey is based on observations made at the Anglo-Australian Telescope. The Sydney-AAO Multi-object Integral field spectrograph (SAMI) was developed jointly by the University of Sydney and the Australian Astronomical Observatory. The SAMI input catalogue is based on data taken from the Sloan Digital Sky Survey, the GAMA Survey and the VST ATLAS Survey. The SAMI Galaxy Survey is funded by the Australian Research Council Centre of Excellence for All-sky Astrophysics (CAASTRO), through project number CE110001020, and other participating institutions. The SAMI Galaxy Survey website is http://sami-survey.org/.\\
 Funding for the Sloan Digital Sky Survey IV has been provided by
the Alfred P. Sloan Foundation, the U.S. Department of Energy Office of
Science, and the Participating Institutions. SDSS-IV acknowledges
support and resources from the Center for High-Performance Computing at
the University of Utah. The SDSS web site is www.sdss.org.

  \bibliographystyle{mnras}
  \bibliography{Voidsbib}
  
  \end{document}